\documentclass[twocolumn]{aastex631}
\usepackage{amsmath, color,booktabs}

\newcommand{\cmmc}{\textsc{\small 21CMMC}}
\newcommand{\cmfst}{\textsc{\small 21cmFAST}}
\newcommand{\cmtool}{\textsc{\small Tools21cm}}
\newcommand{\cmss}{\textsc{\small 21cmSense}}
\newcommand{\pymfs}{\textsc{\small pyMIN}}
\newcommand{\fg}{\textsc{faint galaxies}}
\newcommand{\bg}{\textsc{bright galaxies}}

\newcommand{\hi}{H{~\sc i}}

\graphicspath{{figures/}}

\received{XXX}
\revised{YYY}
\accepted{ZZZ}
\submitjournal{ApJ}

\shorttitle{The 21~cm Minkowsi Functionals}

\shortauthors{Diao et al.}

\begin{document}

\title{Reionization Parameter Inference from 3D Minkowski Functionals of the 21~cm Signals}

\correspondingauthor{Kangning Diao, Yi Mao}
\email{dkn20@mails.tsinghua.edu.cn (KD), ymao@tsinghua.edu.cn (YM)}

\author[0000-0001-7301-2318]{Kangning Diao}
\affiliation{Department of Astronomy, Tsinghua University, Beijing 100084, China}

\author[0000-0002-4965-8239]{Zhaoting Chen}
\affiliation{Jodrell Bank Centre for Astrophysics, School of Physics and Astronomy, The University of Manchester, Manchester M13 9PL, UK}
\affiliation{Institute for Astronomy, The University of Edinburgh, Royal Observatory, Edinburgh EH9 3HJ, UK}


\author[0000-0001-6475-8863]{Xuelei Chen}
\affiliation{Department of Physics, College of Sciences, Northeastern University, Shenyang 110819, China}
\affiliation{National Astronomical Observatories, Chinese Academy of Sciences, Beijing 100101, China}
\affiliation{Key Laboratory of Radio Astronomy and Technology, Chinese Academy of Sciences, Beijing 100101, China}

\author[0000-0002-1301-3893]{Yi Mao}
\affiliation{Department of Astronomy, Tsinghua University, Beijing 100084, China}

\begin{abstract}

The Minkowski Functionals (MFs), a set of topological summary statistics, have emerged as a powerful tool for extracting non-Gaussian information. We investigate the prospect of constraining the reionization parameters using the MFs of the 21~cm brightness temperature field from the epoch of reionization (EoR). 
Realistic effects, including thermal noise, synthesized beam, and foreground avoidance, are applied to the mock observations from the radio interferometric array experiments such as the Hydrogen Epoch of Reionization Array (HERA) and the Square Kilometre Array (SKA). 
We demonstrate that the MFs of the 21~cm signal measured with SKA-Low can be used to distinguish different reionization models, whereas the MF measurement with a HERA-like array cannot be made accurately enough.
We further forecast the accuracies with which the MF measurements can place constraints on reionization parameters, using the standard MCMC analysis for parameter inference based on forward modeling. 
We find that for SKA-Low observation, MFs provide unbiased estimations of the reionization parameters with accuracies comparable to the power spectrum (PS) analysis. Furthermore, joint constraints using both MFs and PS can improve the constraint accuracies by up to $30\%$ compared to those with the PS alone. Nevertheless, the constraint accuracies can be degraded if the EoR window is shrunk with strong foreground avoidance. Our analysis demonstrates the promise of MFs as a set of summary statistics that extract complementary information from the 21~cm EoR field to the two-point statistics, which suggests a strong motivation for incorporating the MFs into the data analysis of future 21~cm observations.

\end{abstract}


\section{Introduction}\label{sec:intro}
The formation and evolution of the very early galaxies is a crucial part of cosmic evolution history. With the unprecedented sensitivity of instruments such as the James Webb Space Telescope (JWST) \citep{2023PASP..135f8001G}, for the first time a large population of very early galaxies at $z\gtrsim 10$ are observed with great details (see, e.g. \citealt{2022ApJ...938L..15C,2023ApJ...946L..13F,2022ApJ...940L..14N,2023MNRAS.518.4755A,2023ApJ...952L...7A}). These observations draw attention to the lack of understanding in early structure formation, as the mass and the luminosity of the early galaxies show potential inconsistencies with predictions from theory \citep[][]{harikane_2022,harikane2022b,harikane2022,bouwens2023}. Observations of the very early Universe can therefore serve as a powerful probe for cosmology and are indispensable to advance the theory of cosmic structure formation.

The formation of early galaxies leads to the cosmic epoch of reionization (EoR; see \citealt{furlanetto_2006} for a review). The baryonic matter in the Universe after recombination mainly consists of neutral hydrogen atoms (\hi), taking up $74\%$ of the mass density of baryons \citep{dodelson2020modern}. After the formation of the luminous objects, ultraviolet (UV) photons are copiously produced, ionizing the \hi\ in the intergalactic medium (IGM).
As a result, \hi\ in the IGM serves as a tracer of the EoR \citep[e.g.][]{Gnedin_1997,Gnedin_2004,furlanetto_2006,Morales_2010,Pritchard_2012}, and is directly linked to the luminosity and spatial distribution of the early galaxies \citep{2022MNRAS.511.3657M}. 

The redshifted 21~cm line, arising from the hyperfine transition of atomic hydrogen \citep{1970ITIM...19..200H}, can be observed by radio telescopes. 
While individual galaxies can not be resolved, high-redshift 21~cm experiments are capable of covering large cosmological volumes to the Gigaparsec scale, ideal for cosmological constraints. Experiments targeting the 21~cm line during EoR include the Low Frequency Array \citep[LOFAR;][]{Harleem_2013,Yatawatta_2013}, Murchison Widefield Array \citep[MWA;][]{tingay_2013}, Precision Array for Probing the Epoch of Reionization  \citep[PAPER;][]{Parsons_2010}, and Hydrogen Epoch of Reionization Array \citep[HERA;][]{DeBoer_2017}. In the future, the low-frequency array of the Square Kilometre Array \citep[SKA;][]{2019arXiv191212699B} will be capable of producing high-fidelity images for EoR measurements \citep{Mellema_2013}.

These 21 cm experiments will likely measure the \hi\ distribution in terms of two-point statistics and several more sophisticated statistical properties extracted from the tomographic 21~cm images.
The 21~cm power spectrum (PS), which is well-studied both in observation \citep[see, e.g.][]{Cheng_2018,Lanman_2019,Trott_2020,theheracollaboration2021results} and in theory \citep[see, e.g.][]{Greig_2015,Li_2019,Pagano_2020}, traces the underlying dark matter distribution at large scales (see, e.g. \citealt{Xu:2019pcz,2022MNRAS.513.5109G}). The signals directly measured by radio interferometers are visibilities in Fourier space \citep{2011A&A...527A.106S}, thereby naturally linked to the PS (e.g. \citealt{Morales_2010}).  
However, the growth of the ionized bubbles induces non-Gaussianity into the 21~cm distribution \citep[see, e.g.][]{Ciardi_2003,Mellema_2006,Shimabukuro_2015,Watkinson_2015,Mondal_2016,Hutter_2020,Shaw_2020} due to reionization patchiness that can be as large as the scale of tens of Megaparsec, as almost no 21~cm signal is emitted from the ionized regions. 
Thus, non-Gaussian statistics are required to give the full description of the clustering in the IGM during reionization.

Limited by the foreground and the principle of interferometry, the 21~cm tomographic data measures the fluctuations, not the total signal --- the positive signals in the data indicate the neutral structures and the negative signals trace the ionized bubbles. Thus the non-Gaussianity affects the clustering of the 21~cm signal through two aspects --- morphology and topology of ionized bubbles. 
Non-Gaussian morphology can be characterized in terms of higher-order correlation functions such as three-point correlation function \citep{Hoffman2019} in configuration space and bispectrum \citep[e.g.][]{Bharadwaj_2004, Yoshiura_2015,Shimabukuro_2017,Hutter_2020,watkinson2021epoch,tiwari2021improving} and trispectrum \citep[e.g.][]{Lewis_2011,Shaw_2019} in Fourier space.
While these statistics have been explored in cosmological theories \citep[e.g.][]{Adshead2013}, these correlation functions do not fully characterize the field distribution in a highly non-Gaussian case \citep{Carron_2011}. 

The size statistics also characterize the morphological information. These include the bubble size distribution \citep[][]{Kakiichi_2017,Giri_2017,Busch_2020,Shimabukuro2021} and neutral island distributions \citep{Xu:2013npa,2017ApJ...844..117X,Giri_2019}, or feature in these distributions such as the largest cluster statistics\citep[][]{pathak2021lcs,lcsobs2023}. Previous works have shown from simulations that the EoR undergoes a phase transition \citep{Furlanett0_2016} during which the ionized bubbles are changed from being small and isolated initially to being connected and percolated as the bubbles expand and merge. 

On the other hand, unlike the conventional Fourier-space polyspectra methods, topological statistics, which are intrinsically defined in configuration space and measured from tomographic data, provide complementary information to the morphology \citep[e.g.][]{Pratten_2012}. 
Topological properties of the field have been characterized in terms of the Minkowski Functionals \citep[MFs; e.g.][]{Gleser_2006,Wang_2015,Yoshiura_2016,Kapahtia_2019,carlo1}, shape finders \citep{bag2019shapefinder} and Betti numbers \citep{Giri_2021}. The EoR can be divided into five stages by quantifying the percolation of ionized bubbles and neutral islands with these statistics \citep{Chen_2019}. 

It is proved and well known that, for an excursion set field, the MFs contain the complete global topological information \citep{Hadwiger_1957}. As such, the MFs are powerful tools for characterizing the topology of cosmological signals. Thus the MFs have been applied to the data analysis in cosmic microwave background (CMB) \citep[e.g.][]{novikov1999minkowski,2017CQGra..34i4002B} and the large-scale structure surveys \citep[e.g.][]{hikage2006primordial,2022ApJ...928..108A,carlo2}.

In this work, we explore the detectability of MFs, taking into account the resolution and instrumental effects of interferometric array experiments. We consider the configurations of a redundant, HERA-like array as well as the SKA. We study the evolution of the observed MFs during different stages of EoR. 
As a demonstration of concept, we extend the \cmmc\footnote{\url{https://github.com/21cmfast/21CMMC}} code \citep{Greig_2015,Greig_2017,Greig_2018} as an MCMC sampler to infer the EoR parameters from the MFs of the 21~cm brightness temperature field. Two reionization models, the \fg \ and \bg \ of \citet{Greig_2017,Greig_2018}, are used to study parameter inferences with MFs in different scenarios.

In previous work, \cite{Kapahtia2021} exploited topological statistics, specifically Contour Minkowski Tensors and Betti numbers in 2D images, to constrain astrophysical parameters from the EoR. In comparison, our work includes observational effects such as the foreground wedge cut to perform a more realistic forecast. In addition, our analysis is not only based on the information in 3D images that contain the information along the LOS but also exploits the MFs as a function of different thresholds as opposed to averaging over threshold values as in \cite{Kapahtia2021}. More modes are therefore included in our analysis. 

The remainder of this paper is organized as follows. In Section \ref{sec:data}, we briefly summarize the theoretical framework to generate mock observations, including the 21~cm brightness field and the instrumental effects. In Section \ref{sec:obs}, we discuss the evolution and underlying information of the observed MFs. We then present the parameter fitting results with the MFs coupled with different instrumental effects in Section~\ref{sec:inf} and make concluding remarks in Section~\ref{sec:sum}. We adopt the values of cosmological parameters in this paper as follows, ($\Omega_\Lambda$, $\Omega_{\rm M}$, $\Omega_b$, $n$, $\sigma_8$, $h$) = (0.69, 0.31, 0.048, 0.97, 0.81, 0.68), in consistency with the {\it Planck} 2015 results \citep{Planck_2016}.

\section{Simulations and Mock Observations}\label{sec:data}

\subsection{Simulations of the 21~cm signals}\label{sec:sim}
The fluctuation of 21~cm emission is represented by its flux density, or equivalently the 21~cm brightness temperature $T_b$. 
In the optically thin approximation, it can be written as \citep[e.g.][]{furlanetto_2006,Mellema_2013}
\begin{equation}
\begin{aligned}
    &\delta T_{b}(\nu) =  27x_{{\rm HI}}(1+\delta_{m})\bigg(1-\frac{T_{\gamma}}{T_{{\rm S}}}\bigg) \\
    &\times \bigg(\frac{1+z}{10}\frac{\Omega_{m}}{0.27}\bigg)^{1/2}\bigg(\frac{\Omega_{b}h}{0.044\times0.7}\bigg)\\
    &\times\bigg(  1 + \frac{1}{a\,H(z)}\frac{\mathrm{d}v_\|}{\mathrm{d} r_\|}\bigg)^{-1}\quad\mathrm{mK},
\end{aligned}
\end{equation}
where $x_{\rm HI}$ is the neutral fraction of the IGM, $\delta_{m}({\mathbf{x}},z) \equiv \rho/\bar{\rho} -1$ is matter density fluctuations, $T_{\gamma}$ is the CMB temperature and $T_{\rm S}$ is the spin temperature. Here we focus on the stage where the gas has been thoroughly heated so that $T_{\rm S} \gg T_{\gamma}$. The signal is also affected by the proper gradient along the line of sight (LOS) of the peculiar velocity, $\mathrm{d}v_\|/\mathrm{d} r_\|$.

\begin{figure*}
    \centering
    \includegraphics[width=\linewidth]{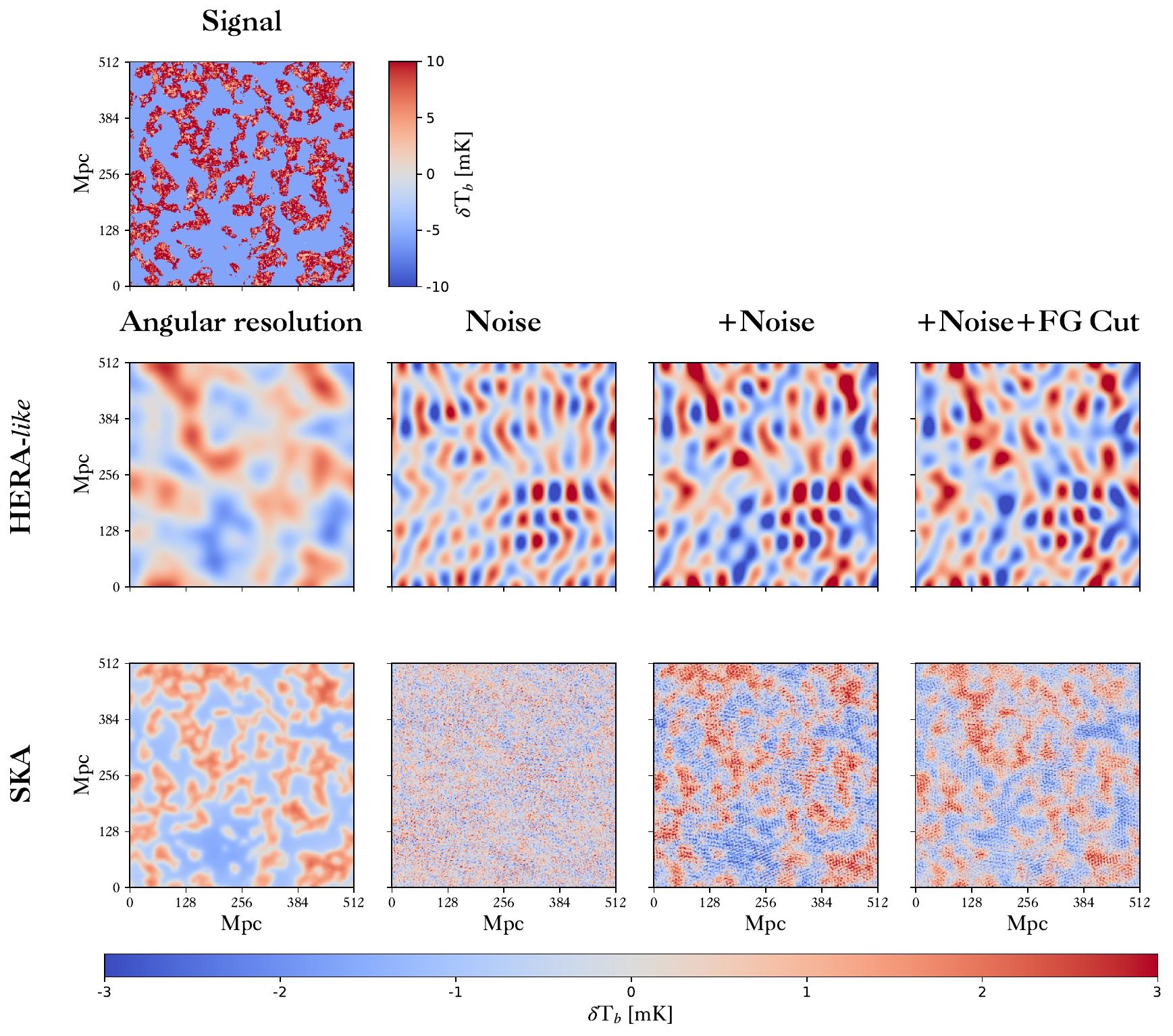}
    \caption{Illustrations of the instrumental effects in mock SKA and HERA-like observations. The top left panel shows a slice of the 21~cm signal at $z=7$ (corresponding to $\bar{x}_{\rm HI}=0.32$) in the \fg\ model. The panels in the middle (bottom) row show the 21~cm signals that are applied with the instrumental effects with the HERA-like (SKA) observations. 
    In each of the middle and bottom rows, panels from the left to the right column show the 21~cm signals that are convolved with the PSF (first column), a realization of thermal noise simulations (second column), the PSF-convolved 21~cm signals with thermal noise applied (third column), the PSF-convolved 21~cm signals that are applied with thermal noise and foreground wedge cut (fourth column), respectively. Note that the colorbar for the middle and bottom panels, as shown at the bottom of the figure, is different from the colorbar for the top left panel.
    }
    \label{fig:Noise}
\end{figure*}

We use the publicly available code \cmfst\footnote{\url{https://github.com/21cmfast/21cmFAST}} \citep{21cmFAST_1.0,21cmFAST_3.0}v3.0.3 to simulate the 21~cm $T_{b}$ field, and the main procedure can be briefly summarized as follows. First, a high-resolution initial condition (IC) is generated through linear perturbation theory. Then the second-order Lagrangian perturbation theory \citep[LPT, e.g.][]{Scoccimarro_1998} is applied to evolve IC to the density and velocity fields at lower redshifts, then smoothing them to lower resolution. Next, following the excursion set theory \citep{Furlanetto_2004}, \cmfst\ estimates the ionization field by comparing the reciprocal of the ionizing efficiency $\zeta$ and the collapse fraction $f_{\rm coll}(\mathbf{x},R,z)$ within a spherical volume. Here, $f_{\rm coll}$ is the fraction of mass collapsed into halos, and $\zeta$ is an aggregation of several physical parameters to describe the photon production rate per baryon in a dark matter halo as described in \cite{21cmFAST_1.0}.

A spherical region around the position $\mathbf{x}$ with the radius $R$ at the redshift $z$ is fully ionized if the condition,
\begin{equation}\label{eqn:ion}
    f_{\rm coll}(\mathbf{x},R,z)\ge \zeta^{-1}\,,
\end{equation}
is satisfied. Following this recipe, \cmfst\ iterates through every cell in the simulation. For each cell, $R$ is initially set to $R_{\rm mfp}$, the mean free path of the ionizing photons or technically the largest possible radius of ionized bubbles. The criterion in Equation~(\ref{eqn:ion}) is then repeatedly checked from large to small radii until a radius $R$ is found to meet the criterion, in which case the central cell of the spherical region is set to be fully ionized. If no radius greater than the cell size $R_{\rm cell}$ meets the criterion, then the cell is partially ionized with the ionized fraction set to be $f_{\rm coll}(\mathbf{x},R_{\rm cell},z)\,\zeta$.

The reionization process can then be simulated by modeling $\zeta$ and $f_{\rm coll}$. In this paper, we focus on a two-parameter model, with the minimum virial temperature ($T_{\rm vir}$) and the ionizing efficiency ($\zeta$). The parameter $T_{\rm vir}$ represents the threshold temperature required for a halo to host ionizing sources, effectively tuning the collapse function $f_{\rm coll}$. 
Higher values of $T_{\rm vir}$ indicate a reduced number of ionizing sources, while larger $\zeta$ values lead to a faster expansion of ionized bubbles. Thus, there exists a degeneracy between $T_{\rm vir}$ and $\zeta$ in terms of their impact on the ionization history. However, it has been noted by \citet{Yoshiura_2015} that despite this degeneracy, the morphology of the ionized regions exhibits distinct characteristics associated with different parameter combinations, as manifested in the MFs.

\begin{table*}
\caption{Telescope parameters at redshift $z=7$ to generate mock observations.}
\begin{center}
\setlength{\tabcolsep}{8mm}{
\begin{tabular}{lcc} \hline
Parameters & HERA-like & SKA-Low\\ \hline
Antenna number&331&512 \\
System temperature ($T_{\rm sys}$) [K] &$100+60\left(\frac{\nu}{300\rm MHz} \right)^{-2.55}$ &$100+60\left(\frac{\nu}{300\rm MHz} \right)^{-2.55}$\\
Dish Area ($A_{\rm Dish}$) [m$^2$]& 153.93& 962.11 \\
Integration time ($t_{\rm int}$) [s] & 10& 10\\
Declination & -38$^{\circ}$ &-30$^{\circ}$\\
Critical frequency ($\nu_{\rm c}$) [MHz] &110 &110\\
Characteristic angular distance($\theta$)&$9.3^{\circ}$ & $6.5^\circ$\\
Frequency resolution ($\Delta \nu$) [MHz] & 0.097 & 0.097\\
Default wedge slope ($m$) & 0.5 & 0.35\\
PSF FWHM [Mpc] & 25 & 3 \\
Averaged thermal noise error\tablenotemark{a} ($\sqrt{\left \langle N^2 \right \rangle}$) [mK] &1.37& 1.01\\
\hline
\end{tabular}}
\end{center} 
\tablenotetext{a}{This is the rms of smoothed thermal noise with 1000 hours observation at $z=7$. The SKA-Low result is consistent with other estimations \citep[e.g.][]{Giri_2018}.}
\label{tab:telescope}
\end{table*}

In this work, all our semi-numerical simulations have a comoving size of $512\times 512 \times 128$ Mpc$^3$ with $\rm (2\, Mpc)^3$ cells. The length of $512$ Mpc in the transverse plane corresponds to an image size of $\sim 3^\circ$, matching the field-of-view (FoV) of the instruments. The length of 128 Mpc along the line-of-sight corresponds to a redshift interval of $\Delta z \sim 0.5$, representing the observations in a frequency sub-band.
In Sections~\ref{sec:data} and \ref{sec:obs}, we adopt a choice of parameters $\zeta = 30$ and $\rm {log}[T_{\rm vir}/K] = 4.70$, corresponding to the \fg\ model in \cite{Greig_2018}, as the benchmark model that shows the characteristics of MFs. The top left panel of Figure~\ref{fig:Noise} shows a slice of simulated brightness temperature at $z=7$. 
In Section~\ref{sec:inf} and afterward, different choices of model parameters using both \fg\ and \bg\ models are discussed. 

\subsection{Mock Observations}\label{sec:noise}
In this section, we use the simulated 21~cm signal to generate mock observations with instrumental effects.  The minimum scales probed are limited by the point spread function (PSF), which for interferometric observation is the synthesized beam of the weighted distribution of interferometric baselines \citep{2016era..book.....C}. The observed signal also has noise, which is assumed to be thermal here. The resulting image will be dominated by the radio foregrounds, which can be avoided by applying a foreground wedge cut \citep{2014PhRvD..90b3018L}.
 
\subsubsection{Thermal Noise}
In an interferometer array, each pair of elements forms a baseline, denoted by its vector components in units of wavelength $(u,v)$, and the interferometric visibility data is recorded independently, contributing to a point in the $uv$ plane. The thermal noise error can be drawn from a Gaussian distribution with the following root-mean-square (rms) for every single baseline \citep[e.g.][]{furlanetto_2006,Giri_2018}
\begin{equation}
     \sigma_N^{\rm single} =\frac{\lambda^2T_{\rm sys}}{\epsilon \Omega A_{\rm Dish}\sqrt{2\Delta \nu t_{\rm int}}},
    \label{eqn:singnoi}
\end{equation}
where $\Omega\approx \theta^2$ is the beam area, $\lambda$ is the signal wavelength, $T_{\rm sys}$ is the system temperature of the telescope, $\Delta \nu$ is the frequency resolution, $t_{\rm int}$ is the integration time, $A_{\rm Dish}$ is the physical area of station and $\epsilon$ is an efficiency factor which follows
\begin{align}
\begin{split}
\epsilon= \left \{
\begin{array}{ll}
    1,   &\nu<\nu_{\rm c}\\
   \left ( \frac{\nu_{\rm c}}{\nu}\right )^2, & \nu\ge\nu_{\rm c}.
\end{array}
\right.
\end{split}
\eqnum{}
\end{align}

When calculating the thermal noise, we discretize the $uv$ space into 2D grids. Assuming $N_{\rm uv}$ baselines fall in the same pixel in one night, and the final visibility data is the coherent average of $N_{\rm d}$ nights, then the noise rms of this pixel is calculated as
\begin{equation}
 \sigma_N^{\rm pixel} =\frac{ \sigma_N^{\rm single}}{\sqrt{N_{\rm d}N_{uv}}},
\end{equation}
where we assume that natural weighting is applied when producing the image cube. Throughout this paper, we use natural weighting, i.e. the grid weights are the number of baselines in each $u$-$v$ grid.

Here we consider two telescope configurations, 
HERA-like and SKA-Low, whose parameters are summarized in Table \ref{tab:telescope}. We calculate the SKA-Low thermal noise through the publicly available code \cmtool\footnote{\url{https://github.com/sambit-giri/tools21cm}} \citep{Giri_2020}, following the procedure in \citet{Giri_2018} and \citet{Hassan_2019}. We generate the HERA-like thermal noise in the same manner, based on \cmtool  \ and \cmss\footnote{\url{https://github.com/steven-murray/21cmSense}} \citep{Pober_2013,Pober_2014}. We refer to the latter settings as ``HERA-like'' for reasons discussed later in Section \ref{sec:res}. Slices of the simulated 1000 hours HERA-like and SKA-Low noise are shown in Figure \ref{fig:Noise}.

\subsubsection{Point Spread Function}
\label{sec:res}
The interferometer measures the Fourier modes of the sky signal. The visibility data in the $u$-$v$ space are averaged into the $u$-$v$ grids and then transformed back to the angular space for imaging. Since the gridded visibility is the multiplication of the sky signal in Fourier space with the grid weights, the images are then the convolution of the sky signal and the synthesized beam, which is the Fourier transform of the grid weights \citep{2016era..book.....C}. The synthesized beam is also referred to as the point spread function (PSF). In the case of natural weighting, the PSF is the renormalized Fourier transform of the baseline distribution.

\begin{figure}
    \centering
    \includegraphics[width=\linewidth]{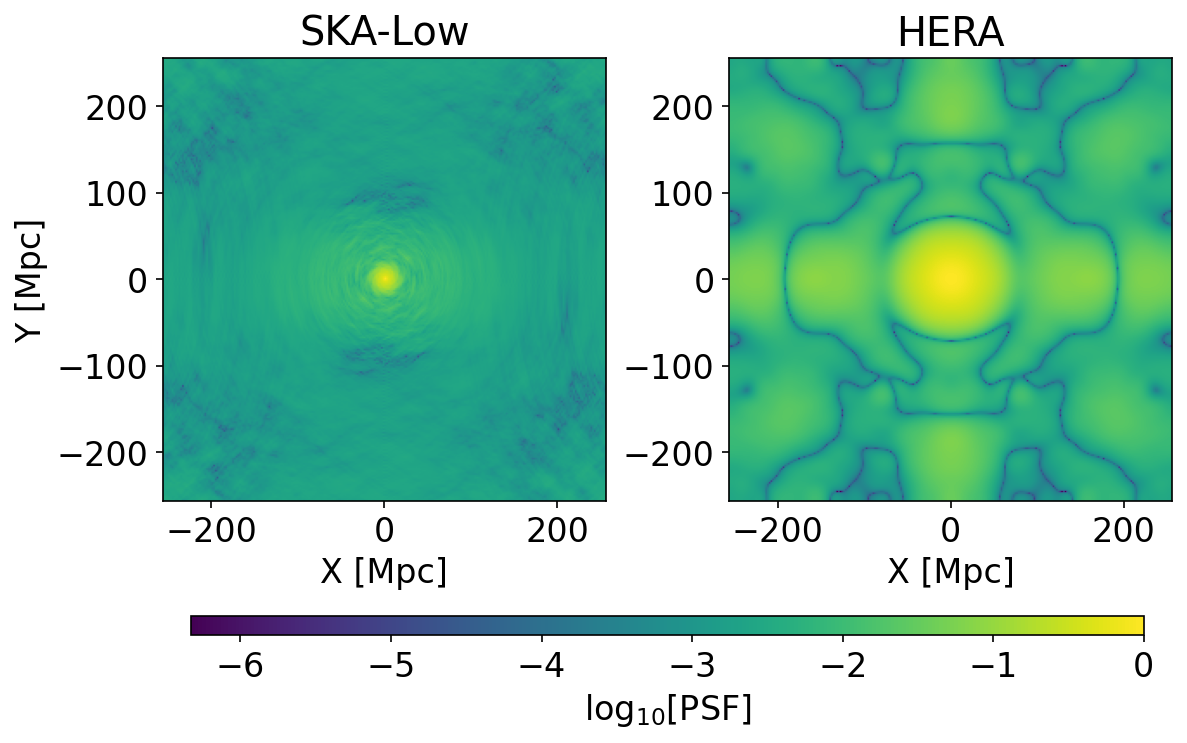}
    \caption{The PSFs used for the mock observations at $z=8$ using SKA-Low (left) and for HERA-like observations (right), respectively.}
    \label{fig:psf}
\end{figure}

For SKA-Low, we use the v3 station layout \citep{2020arXiv200312744D} to simulate the baseline distribution of a typical tracking observation. We choose the phase center to be the EoR0 field \citep{2021PASA...38...57L} at ${\rm RA}=0\,{\rm h}$, ${\rm Dec} =-30\,{\rm deg}$. The observation time of one track is assumed to be a total of 4 hours with a time resolution of 10 seconds. The coherent average of multiple nights is assumed to reach the total integration time of 1000 hours. The baseline distribution is simulated using the \textsc{OSKAR}\footnote{\url{https://github.com/OxfordSKA/OSKAR}} code \citep{5613289}. The baseline distribution is then gridded into $256\times 256$ u-v grids, with the cell size corresponding to the 512\,Mpc box size in the transverse plane. The baseline distribution is then Fourier transformed to obtain the PSF. The PSF at $z=8$ is shown in the left panel of Figure \ref{fig:psf} for the purpose of illustration, with the corresponding spatial resolution of the PSF listed in Table \ref{tab:telescope}. 

HERA, on the other hand, is a redundant baseline array designed for redundant calibration and power spectrum estimation \citep{2020MNRAS.499.5840D}. Here, we do not intend to simulate realistic drift scan observations of HERA, but to use its array layout to generate PSFs that are qualitatively different from SKA-Low observations for comparison. Throughout this paper, we refer to this scenario as ``HERA-like'' observations. We use the \textsc{hera\_sim}\footnote{\url{https://github.com/HERA-Team/hera_sim}} code to retrieve the station layout to calculate the baseline coordinates. We then symmetrize the baseline distribution, as a crude approximation of a coherent average of different nights when the same patch of the sky passes through. The baseline distribution is then gridded to obtain the PSF. The resulting PSF at $z=8$ is shown in the right panel of Figure \ref{fig:psf}.

The simulated PSFs are then convolved with the signal cubes. Note that we do not include primary beam attenuation in generating the mock observation, since the $512$ Mpc box length is within the primary beam of the instruments. In our case, the primary beam attenuation is simply multiplied in the signal, which is then divided out. This step only increases the thermal noise level of the pixels. Therefore, we note to the readers that the thermal noise simulated can only be achieved with an integration time larger than 1000 hours in real observations. The chromaticity of the primary beam may introduce additional structure in the observed 21~cm signal, whose effects are beyond the scope of this work.


\subsubsection{Foreground Wedge}
\label{sec:fgwedge}
The major challenge for the EoR 21~cm line observation is the contamination of the foregrounds, which is four to five orders of magnitude higher than the signal. We follow the foreground avoidance method by filtering out the foreground-dominated $k$-modes outside the EoR window. A thorough explanation of the EoR window can be found in e.g. \citet{Liu2020}, and we briefly summarize the main idea below.

In the cylindrical 2D Fourier (i.e.\ $k_{\parallel}$-$k_{\perp}$) plane, where $k_{\parallel}$ ($k_{\perp}$) is the comoving wave number parallel (perpendicular) to the LOS, respectively, the region satisfying $k_{\parallel}<mk_{\perp}$ is dominated by the foregrounds, but the regions where $k_{\parallel}>mk_{\perp}$ (i.e.\ ``EoR window'') is believed to have negligible foreground contamination so it can be used to study the 21~cm signal. Here \textit{m} is the so-called ``foreground wedge'' slope 
\begin{equation}
    m=\frac{DH_{0}E(z)\sin \theta}{c(1+z)},
\end{equation}
where $H_{0}$ is the Hubble constant, $D$ is the comoving distance, $E(z) = H(z)/H_{0}$, the ratio of Hubble parameter at $z$ and Hubble constant at present, 
and $\theta$ is the characteristic angular distance outside of which the foreground emission is negligible due to power beam attenuation.
In the optimistic case, the foreground emission is contained within the primary beam FoV of the instrument. Our estimation of the characteristic angular distance $\theta$ for SKA, as given in Table~\ref{tab:telescope}, is derived from \citet{2019arXiv191212699B}, with the interpolation to our desired frequency using the relation \(\theta \propto 1/\nu\). Specifically, the value of \(\theta\) herein is twice the FoV presented in \citet{2019arXiv191212699B}, consistent with \citet{2023MNRAS.524.3724C}. The wedge with the value of \(\theta\) herein is on an optimistic estimation \citep{2023MNRAS.524.3724C}. For the HERA-like telescope, the value of $\theta$ is estimated using \cmss.


It is worth noting that, the feature of foreground wedge arises from the chromaticity of the baselines, i.e. the fact that the angular scale measured by a baseline changes with frequency (see e.g. Figure 5 of \citealt{2018ApJ...869...25M}). For tomographic imaging where the visibilities are averaged in a map-making process, the chromaticity no longer exists and there is no well-defined foreground wedge (see e.g. \citealt{2021MNRAS.500.2264H}). However, the residual foreground cleaning process is expected to remove the foreground power at high $k_\parallel$ and effectively recover an observation window above the wedge (see e.g. \citealt{2023MNRAS.524.3724C}). Therefore, we expect a Fourier filter based on the foreground wedge criterion can be used to extract the \hi\ signal in the images. In practice, we Fourier transform the 21~cm signal from real-space images to the \(k\)-space using the 3D FFT, flag all \(k\) modes with \(k_{\parallel} < mk_{\perp}\) as zero, and then Fourier transform the signal back to obtain the real-space images with foreground wedge cut.

For EoR observations, telescopes with wide FoV are typically used. For instruments like HERA and SKA-Low, the beam sidelobes are non-negligible up to the physical horizon. The chromaticity of the sidelobes introduces foreground contamination at high $k_\parallel$ modes, lifting the slope of the foreground wedge \citep{2016ApJ...819....8P}. The sky signal is cut off at the physical horizon, and therefore there is a large wide-field effect around the horizon line $\sin \theta = 1$ \citep{2015ApJ...804...14T}. These effects need to be carefully studied and mitigated in real observations and are beyond the scope of our work. We assume the optimistic scenario where the foreground wedge is suppressed to the angular scale of the primary beam with the values listed in Table \ref{tab:telescope}. Larger values of $m$ are also explored later to examine the impact of the position of the foreground wedge on MFs in Section \ref{sec:mfsconf}. Finally, we note that identification of ionized regions in the presence of foregrounds may be achieved through machine learning techniques for future SKA-Low observations \citep{2024MNRAS.tmp..283B}, which we leave for future work.

\section{Minkowski Functionals of the 21~cm Brightness Temperature Field}\label{sec:obs}
\subsection{Definition of Minkowski Functionals}

\begin{figure*}[htbp]
    \centering
    \includegraphics[width=\linewidth]{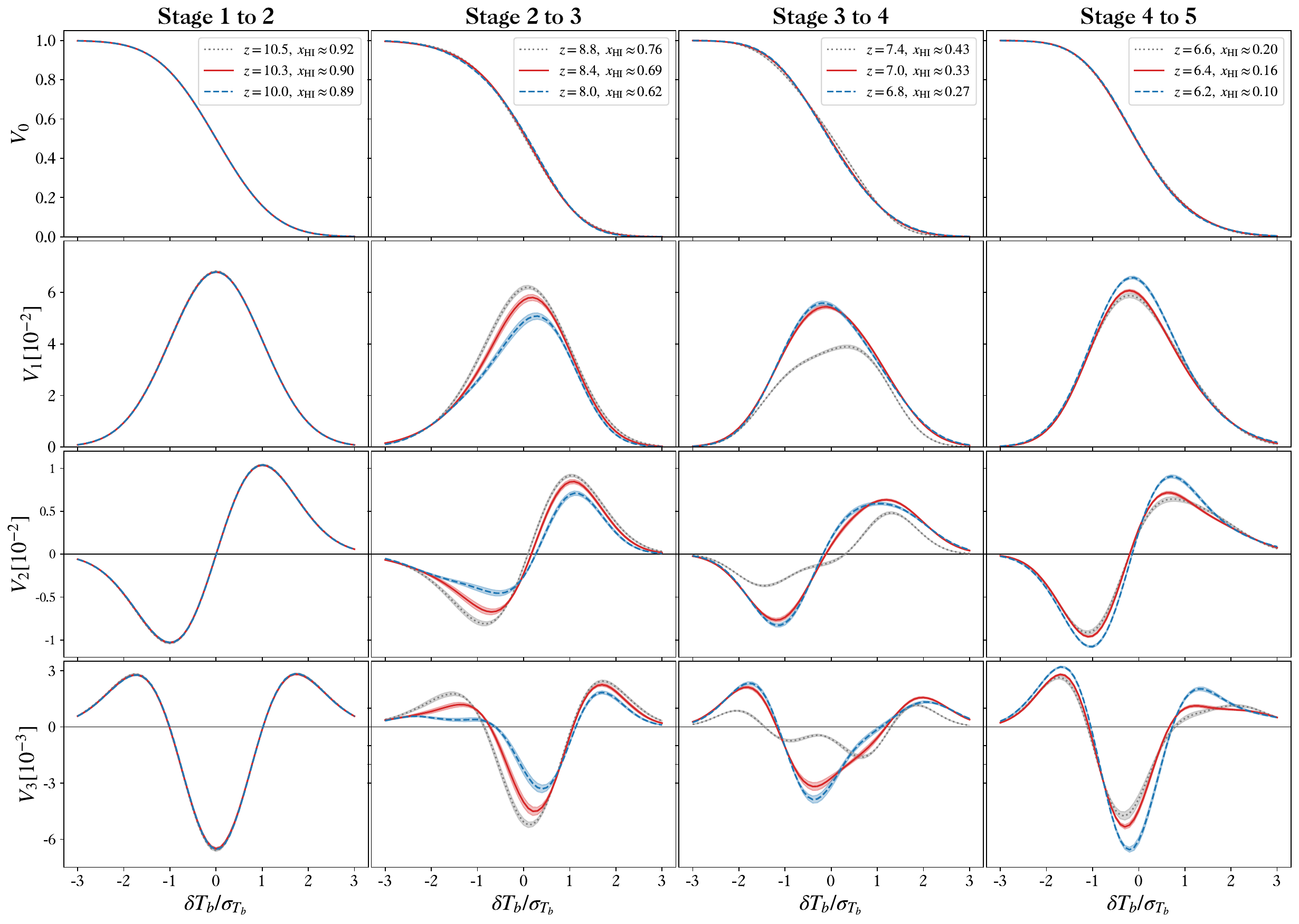}
    \caption{The MFs around transition points with SKA-Low noise between different stages. We show four MFs (from top to bottom) -- $V_{0}$, $V_{1}$,$V_{2}$, and $V_{3}$, respectively, around four different transition points (from left to right). The grey-dotted, red solid, and blue dashed lines represent the average MFs before, during, and after the transition point. The 1$\sigma$ errors of each line are shown in the shaded region with the same color.}
    \label{fig:MFsEvo}
\end{figure*}

The MFs \citep{Minkowski_1903} are a set of functionals describing the morphology of contour surfaces of a field. In three-dimensional space, MFs consist of four independent functionals \citep{Schmalzing_1997}, which are defined as follows:
\begin{align}
    V_{0}(\nu)&=\frac{1}{V}\int_{V}\mathrm{d}^3x\ \Theta[u(\boldsymbol{x})-\nu\sigma]\\
    V_{1}(\nu)&=\frac{1}{6V}\int_{\partial F_{\nu}}\mathrm{d}s\\
    V_{2}(\nu)&=\frac{1}{6\pi V}\int_{\partial F_{\nu}}\mathrm{d}s\ [\kappa_1(\boldsymbol{x})+\kappa_2(\boldsymbol{x})]\\
    V_{3}(\nu)&=\frac{1}{4\pi V}\int_{\partial F_{\nu}}\mathrm{d}s\ \kappa_1(\boldsymbol{x})\kappa_2(\boldsymbol{x})
\end{align}
where $u(\boldsymbol{x})$ is the field being studied, $\sigma$ is the rms of the field, $\nu\equiv u_{\rm thr}/\sigma$ represents the relative threshold of the field, $V$ is the volume that contains the field, $\Theta$ represents the Heaviside step function, $\partial F_{\nu}$ denotes the contour surface whose threshold value is $u_{\rm thr}$, and $\kappa_{1}$ and $\kappa_{2}$ are the principle curvatures of the $\partial F_{\nu}$. The morphological meanings of $\{V_{0},V_1,V_2,V_{3}\}$ are the volume inside, surface area, mean curvature and the Euler characteristics of $\partial F_{\nu}$. Here, ``inside'' denotes the region where $u(\boldsymbol{x})>u_{\rm thr}$. The $V_{3}$ is related to the genus number $g$ by $g=1-V_{3}/2$, which describes the homotopy of the field by $gV=(\rm number\ of\ 
holes)-(\rm number\ of\ isolated\ islands)+1$.

We take the Kodenderink method \citep{Koenderink_1984} to compute MFs. In this method, $\kappa_1+\kappa_2$ and $\kappa_1 \kappa_2$ are represented by the first- and second-order partial derivatives of the field, whereby these integrals can be transformed into volume integrals. We developed a Python package \pymfs\footnote{\url{https://github.com/dkn16/pyMIN}} to perform the calculation. Unless otherwise mentioned, we calculate the MFs from $-3\sigma$ to $3\sigma$ with a sampling interval of $0.1\sigma$, thus including 61 data points in total for each functional.

A special case of the MFs is that of the Gaussian random field, which can be analytically expressed as follows. 
\begin{align}
\label{eqn:ana1}
    V_{0}(\nu)&=\frac{1}{2}-\frac{1}{2}\Phi\left(\frac{\nu}{\sqrt{2}}\right)\\
    V_{1}(\nu)&=\frac{2}{3}\frac{\lambda}{\sqrt{2\pi}}e^{-\nu^2/2}\\
    V_{2}(\nu)&=\frac{2}{3}\frac{\lambda^2}{\sqrt{2\pi}}\nu e^{-\nu^2/2}\\
    V_{3}(\nu)&=\frac{\lambda^3}{\sqrt{2\pi}}(\nu^2-1) e^{-\nu^2/2}
    \label{eqn:ana4}
\end{align}
where 
$\Phi$ is the Gaussian error function, $\lambda\equiv\sqrt{\sigma_1/6\pi\sigma}$ and $\sigma_1$ is the rms of the $|\nabla u|$ field. These functions provide a template for comparison. In particular, MFs closer to the Gaussian analytical form represent a more Gaussian field. 

Throughout this paper, the 21~cm brightness temperature field is normalized by its standard deviation before calculating the MFs, a different treatment from that of \citet{Chen_2019} where the MFs are calculated directly on the temperatures. As we focus on mock observations with different levels of thermal noise and reionization models, the normalization procedure sets the comparison between different scenarios on a fairer basis. 

\subsection{Evolution during Reionization}
\label{sec:mfsevo}


We begin by investigating the evolution of MFs during the reionization process through simulated mock observations. In our study, we employ a fiducial observation scenario involving a 1000h integration time. In this section, we assume no foregrounds as the most optimistic scenario and assume the configuration of SKA-Low to generate the mock observations, but leave the discussion of the impact of filtering out foreground wedge to Section~\ref{sec:mfsconf}.

To investigate the evolution of MFs, we examine the critical redshifts before, during, and after each of the four transition points described in \citet{Chen_2019}. The average neutral fractions of the IGM at the transition points are $\bar{x}_{{\rm HI}} \approx 0.9$, 0.7, 0.3, and 0.15. For detailed discussions of the different stages of EoR and the transitions, see \citet{Chen_2019}. Here we recap the main features of these five stages as follows. For every redshift, we construct ten independent mock observations and calculate the corresponding mean and the sampling variance of the MFs. 
\begin{itemize}
    \item Ionized bubbles stage (\(\bar{x}_{{\rm HI}} > 0.9\)): a small number of isolated ionized bubbles begin to appear at the high-density peaks within the volume.
    \item Ionized fibers stage (\(0.7 < \bar{x}_{{\rm HI}} < 0.9\)): an increasing number of ionized bubbles appear and connect, forming a large, fiber-like structure that extends throughout the entire simulation box.
    \item Sponge stage (\(0.3 < \bar{x}_{{\rm HI}} < 0.7\)): most ionized regions become interconnected, with neutral and ionized regions tangling up to form a sponge-like complex structure.
    \item Neutral fibers stage (\(0.15 \lesssim \bar{x}_{{\rm HI}} < 0.3\)): neutral regions contract to form large fibers embedded within the ionized regions.
    \item Neutral islands stage (\(\bar{x}_{{\rm HI}} \lesssim 0.15\)): the remaining large neutral regions become isolated islands, which continue to shrink and eventually disappear.
\end{itemize}

To untangle the effects of noise and PSF, we start by removing the thermal noise component in the mock observation and investigate the noise-free signal box convolved with the PSF. We find that the noise-free scenario exhibits consistent characteristics as in \citet{Chen_2019}, as the configuration of SKA-Low is capable of resolving the small-scale structure of the 21~cm image shown in the bottom left panel of Figure \ref{fig:Noise}.
To avoid redundancy, we choose not to present our result of MFs in the noise-free scenario and refer the interested readers to Figure~2 of \citet{Chen_2019}. We then proceed to calculate the MFs convolved with PSF and assume the SKA-Low thermal noise with 1000h observation time, and show the results in Figure \ref{fig:MFsEvo}. Each line in the figure represents the mean MFs along with the shaded regions being their $1\sigma$ errors, derived from ten independent realizations. 
The presence of noise impedes a direct correlation between the evolution of MFs and that of ionized bubbles.
Despite the presence of biases in all MFs, the relatively small $1\sigma$ errors indicate that the noised MFs is dominated by the 21~cm signal. Consequently, these biased MFs still exhibit distinct features 
that trace the evolution of the EoR which we summarize as follows.


During the transition from the ionized bubble stage (Stage 1) to the ionized fiber stage (Stage 2) ($\bar{x}_{\rm HI}\sim 0.90$), 
the reionization process remains largely Gaussian, corresponding to relatively small bubbles of ionized regions. As the temperature field is normalized, MFs at $\bar{x}_{\rm HI}\gtrsim 0.90$ are almost identical. 
The application of PSF smoothing and the presence of Gaussian thermal noise obscures the subtle non-Gaussian signatures that start to emerge at $\bar{x}_{\rm HI}\sim 0.90$, originating from the growth of the ionized bubbles. We conclude that the first transition point of morphology described by the MFs can not be observed in realistic observations.


As the reionization process progresses, distinct features begin to emerge within the noised MFs. The transition from the ionized fiber stage to the sponge stage (Stage 2 to Stage 3, $\bar{x}_{\rm HI}\sim 0.70$) coincides with the emergence of large connected ionized regions. The MFs deviate from the analytical Gaussian form and start to evolve with the evolution of neutral fraction, demonstrating the non-Gaussianity of the EoR. Distinctive features of the peaks and troughs in $V_{1,2,3}$ start to evolve. The merging of ionized bubbles decreases the surface area of the ionizing front, as shown in the decreasing of the peak in $V_1$ around $\delta T_b \sim 0$.
Similarly, a decrease of peak structures also appears in $V_2$ and $V_3$. From $\bar{x}_{\rm HI}\sim 0.7$ onwards, the non-Gaussiantiy of the EoR starts to dominate, which makes the MFs a powerful probe of its evolution.


During the transition from the sponge stage to the neutral fiber stage (Stage 3 to Stage 4, $\bar{x}_{\rm HI}\sim 0.32$), the originally decreasing peaks of MFs start to increase as seen in the third column of Figure \ref{fig:MFsEvo}. It indicates that the dominating morphology is no longer large neutral regions surrounding the ionized bubbles and filament structures. Instead, the neutral IGM is enclosed by the ionized region, forming small patches of neutral islands \citep{2017ApJ...844..117X} as reionization progresses. This shift in morphology can also be seen by the fact that the position of the $\delta T_b \sim 0$ peak in $V_1$ and trough in $V_3$ shift from positive to negative values. The MFs trace closely the evolution of the neutral islands, and the morphology of the islands can be used to indicate the evolution of IGM towards the end of EoR \citep{2019MNRAS.489.1590G}.



During the transition from the neutral fiber stage to the neutral island stage (Stage 4 to Stage 5, $\bar{x}_{\rm HI}\sim 0.16$), the majority of the IGM become ionized, and the features of MFs are predominantly influenced by thermal noise, mirroring the ionized bubbles at $\bar{x}_{\rm HI}\gtrsim 0.9$. However, as seen in the fourth column of Figure \ref{fig:MFsEvo}, the MFs still demonstrate sizeable non-Gaussian features. The neutral islands do not trace the luminous objects and their host clusters, and therefore they are much more non-Gaussian. The peak features in $V_{1,2,3}$ continue to grow, indicating that the islands are increasingly disconnected as the Universe gets more ionized.


In conclusion, the evolution of the noised MFs shows mixed effects of Gaussian smoothing from the thermal noise, as well as evolution in its features tracing the morphology of EoR. For deep observations using SKA-Low, the signal-to-noise level required for resolving the morphology can be reached and MFs can be used to constrain the reionization, which we will discuss in Section~\ref{sec:inf}. 


\subsection{The Impact of Foreground Wedge}
\label{sec:mfsconf}


\begin{figure}
    \centering
    \includegraphics[width=\linewidth]{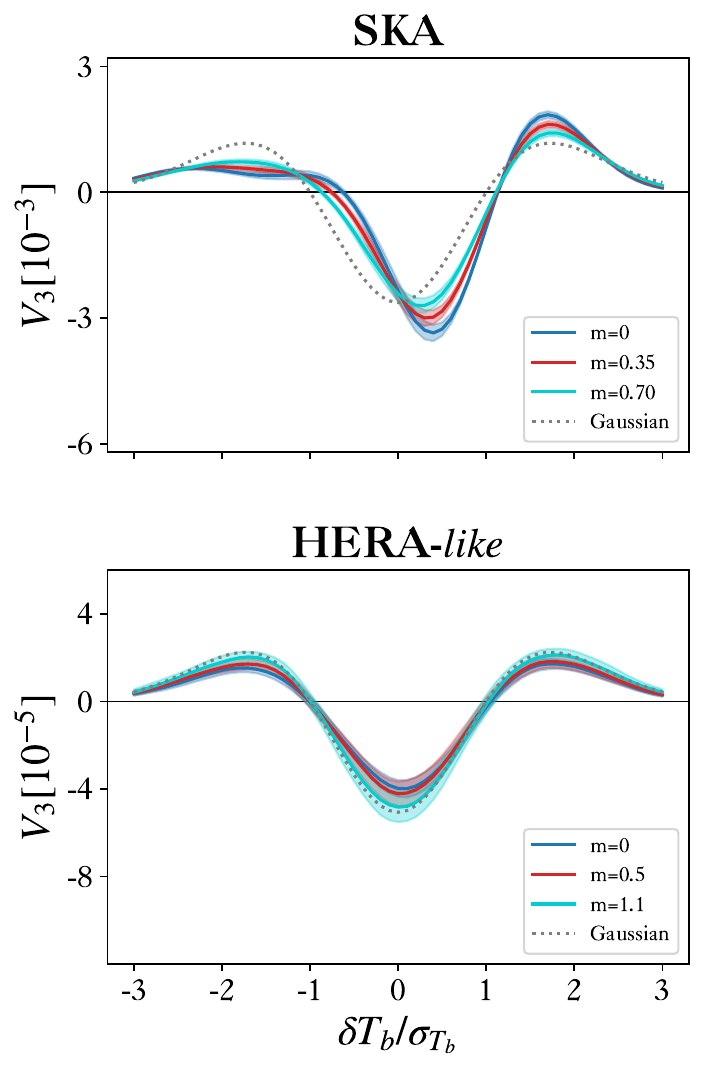}
    \caption{The MF $V_{3}$ at $z=8$, assuming thermal noise with SKA-Low deep (top) and HERA-like (bottom) observations, respectively. Here we assume different scenarios of foreground cut --- the no-foreground-cut scenario (blue solid line), an optimistic (default) foreground wedge slope $m=0.35$ ($0.5$) for SKA (HERA-like) observations  (red solid line), and a more realistic scenario where the value of the wedge slope $m$ is doubled from the default value (cyan solid line).  All fields are normalized to have the same Gaussian template (grey dotted line). We also show the 1$\sigma$ errors in shaded regions with the same color.}
    \label{fig:MFsConf}
\end{figure}



The continuum foreground emission dominates the signals on large scales along the LOS, thereby causing the loss of large-scale information. In this subsection, we explore the impact of different observational configurations on the observed MFs and test the robustness of MFs in the presence of foregrounds. As mentioned in Section \ref{sec:data}, we focus primarily on SKA-Low, while only presenting a HERA-like case for extreme comparisons. 
 



To avoid redundancy, in this subsection, we only show the comparisons of $V_3$ in different settings of observations. The Euler characteristic $V_3$ is related to the genus number of the EoR field, a well-studied morphological indicator for EoR (e.g. \citealt{2008ApJ...675....8L}). The differences in $V_3$ and their implications can be trivially extended to other MFs.

In Figure \ref{fig:MFsConf}, we show the results of the MF $V_{3}$ at $z=8$ as an illustrative example to highlight the characteristics of different scenarios. For a fair comparison, we normalize each field with the standard deviation of the $m=0$ case instead of their respective standard deviations. The Gaussian template with the unit standard deviation is also shown to demonstrate the (non)-Gaussianity of $V_3$ in different scenarios, which we discuss in detail below.

\textbf{(1) No Foreground Cut Case.} --- We begin by considering the most optimistic scenario where foreground contamination can be completely removed, as indicated by the ``$m=0$'' case in Figure \ref{fig:MFsConf}. In this case, the $V_{3}$ obtained from SKA-Low exhibits notable deviations from the Gaussian template, 
while $V_3$ with HERA-like observations is Gaussian-like.

Despite the resemblance with a Gaussian field, 
the amplitude of $V_3$ is attenuated and the $1\sigma$ region is extended compared to the Gaussian case, both of which arise as the consequences of PSF smoothing. 
For HERA-like redundant layouts, the sky is only sampled at some fixed angular scales, with the majority of the morphological information lost.
Consequently, with less remaining information, the field becomes more susceptible to thermal noise fluctuations, leading to larger error bars.

\textbf{(2) Foreground Effects.} ---  
We consider two scenarios of foreground contamination.
The first scenario --- $m=0.35$ ($0.5$) for SKA (HERA-like) observations, respectively --- represents an optimistic case where foreground removal algorithms successfully mitigate the foreground, bringing the wedge down to the primary beam FoV as listed in Table \ref{tab:telescope}. The second scenario represents a more realistic case where the foreground slope $m$ is magnified by a factor of 2 due to incomplete foreground removal in beam sidelobes. 

Across all cases with foregrounds, we observe that the loss of modes leads to a decline in non-Gaussianity compared to the no-foreground-cut case. For the SKA-Low configuration,
the presence of the foregrounds results in fewer features in the MFs.
As expected, raising the foreground wedge results in more Gaussian MFs with less information.
Regarding the HERA-like configuration, we observe an increase in the noise level and the peak structure, due to the loss of information as well as the numerical artifacts from filtering out the $k_\parallel$ modes.

In conclusion, we find that for SKA-Low observations, the non-Gaussian features of MFs are robust against the presence of foregrounds, and can be observed assuming a realistic level of foreground mitigation.

\section{Reionization Parameter Inference with MFs}
\label{sec:inf}
\subsection{The Impact of Reionization Parameters}
\label{sec:mfsparams}
\begin{figure*}[htb!]
    \centering
    \includegraphics[width=\linewidth]{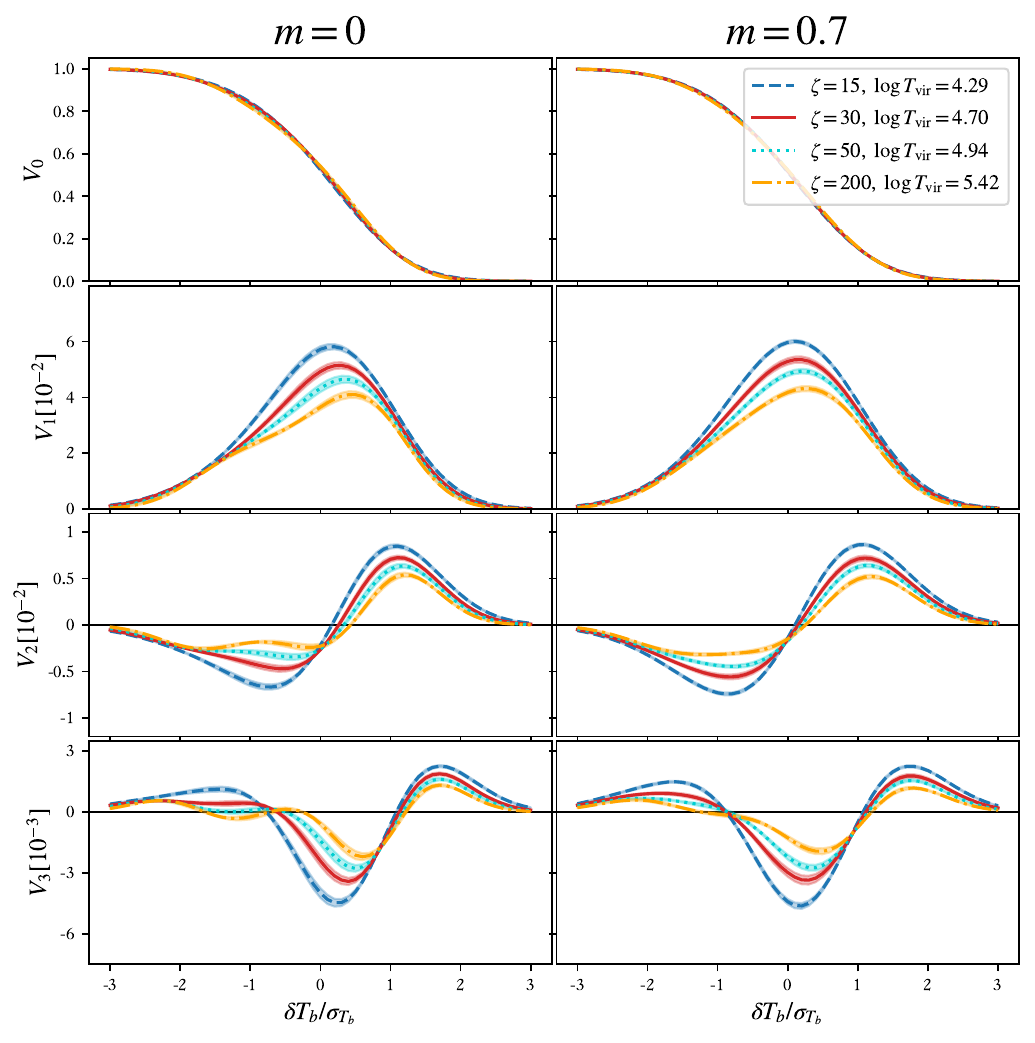}
    \caption{The average MFs and corresponding $1\sigma$ errors in shaded regions, assuming SKA-Low noise. We consider the no foreground cut scenario (left) and the foreground scenario with the wedge slope $m=0.7$ (right), respectively. We show the results by varying four different sets of reionization parameters that all yield $\bar{x}_{\rm HI}\approx 0.60$ at $z=8$.}
    \label{fig:MFsPar}
\end{figure*}

In this section, we investigate the impact of reionization parameters on the observed MFs of the 21~cm signal and in turn, use the MFs to constrain the EoR.
For each scenario, we choose four sets of reionization parameters that yield identical reionization history, with $\left( \zeta, {\rm log_{10}[{T_{vir}/K}]}\right) = \left(15,4.29\right)$, $\left(30,4.70\right),\left(50,4.94\right)$ and $\left(200,5.42\right)$. As discussed in Section \ref{sec:data}, increasing values of $\left( \zeta, {\rm log_{10}[{T_{vir}/K}]}\right)$ lead to more inhomogeneous reionization, as the reionization sources will be less numerous. We generate ten independent mock realizations for each parameter set and compute the corresponding MFs. For simplicity, we show the results at $z=8$ with $\bar{x}_{\rm HI}=0.60$, and the conclusions can be trivially extended to other redshifts. For simplicity, in the rest of this paper, we use $\rm {log}_{10}T_{vir}$ to denote ${\rm log_{10}[{T_{vir}/K}]}$.

The left panel of Figure \ref{fig:MFsPar} shows the MFs corresponding to different parameter values, assuming SKA-Low noise and no residual foreground. Despite the presence of noise, the MFs successfully break the degeneracy among the parameter sets, as evidenced by the varying shapes of the MFs in four cases.
Specifically, the MFs $\{V_1,V_2,V_3\}$ show substantial variations in dependence on the parameter values, while $V_0$ exhibits negligible changes. One prominent feature is that higher values of $\zeta$ and $T_{\rm vir}$ tend to produce lower peaks in $\{V_1,V_2,V_3\}$. This phenomenon can be attributed to the ``faint'' model, characterized by low $\zeta$ and $T_{\rm vir}$, which generates numerous small ionized bubbles. Although some of these small bubbles may be smoothed out due to limited resolution and thermal noise, a substantial number remains in the signal. In contrast, the ``bright'' model, characterized by higher $\zeta$ and $T_{\rm vir}$, leads to larger but fewer ionized bubbles with less structure. 
Therefore, it can be seen that the MFs are more sensitive to the reionization parameters when the reionization model yields a larger population of small ionized bubbles. 
To further test the robustness of MFs in distinguishing reionization models with the foreground cut, we choose $m=0.7$,
which represents a case of severe foreground contamination, and calculate the MFs in the four sets of parameters as shown in the right panel of Figure~\ref{fig:MFsPar}. 
As expected, the $\{V_0,V_1,V_2,V_3\}$ curves in this case appear slightly more Gaussian compared to the no-foreground-cut scenario. Nevertheless, the significant differences of the MFs $\{V_1,V_2,V_3\}$ between different parameter sets still provide strong evidence for the robust constraining power of MFs.

In conclusion, MFs exhibit considerable power in distinguishing different reionization models, even in the presence of instrumental effects and foreground cuts. This indicates that the MFs are likely to be exploited in constraining reionization parameters --- a topic we will discuss in the remainder of this section.

\subsection{\cmmc \ Setup}
Now we quantify the constraining power of MFs. For the forecast study, we extend the \cmmc\ package \citep{Greig_2015} and use MFs as the summary statistics for constraining EoR. We describe the procedure below.
\begin{figure*}[htbp]
    \centering
    \includegraphics[width=0.496\textwidth]{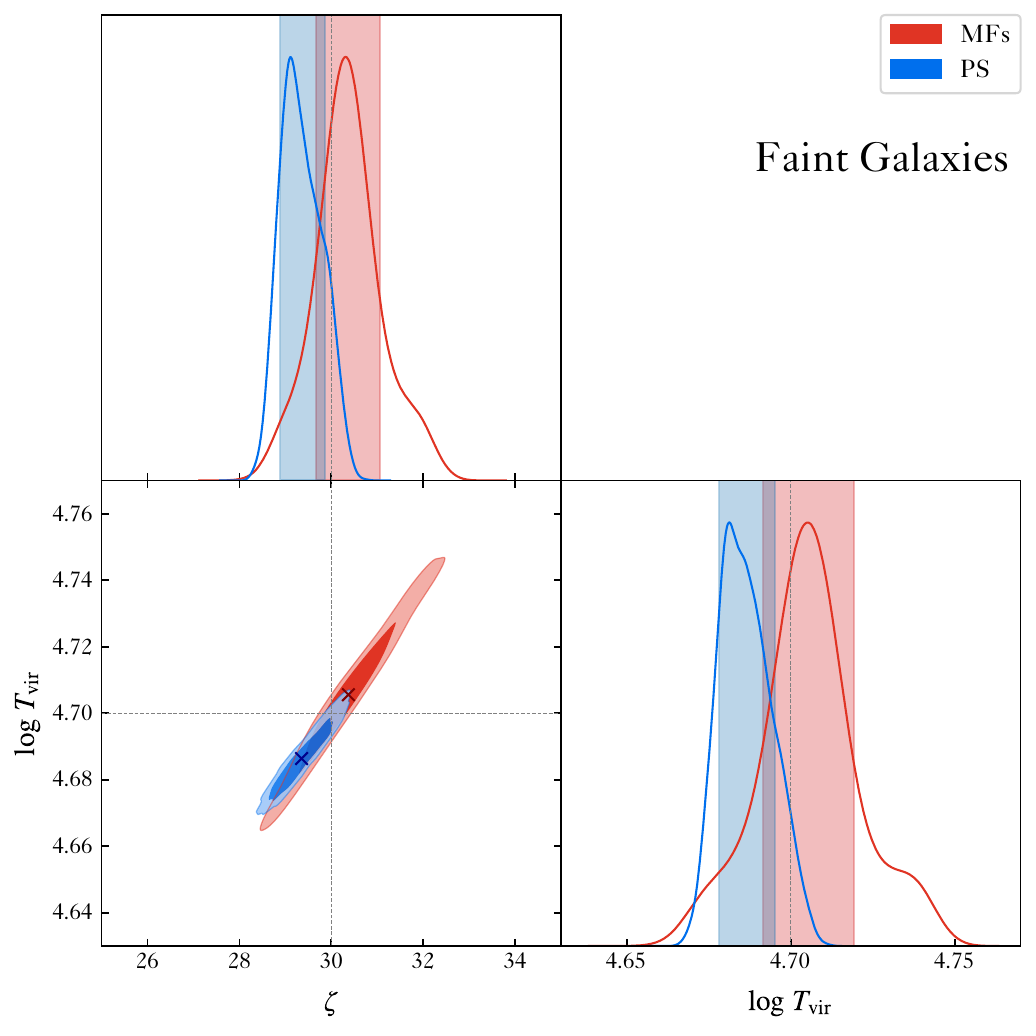}
    \includegraphics[width=0.496\textwidth]{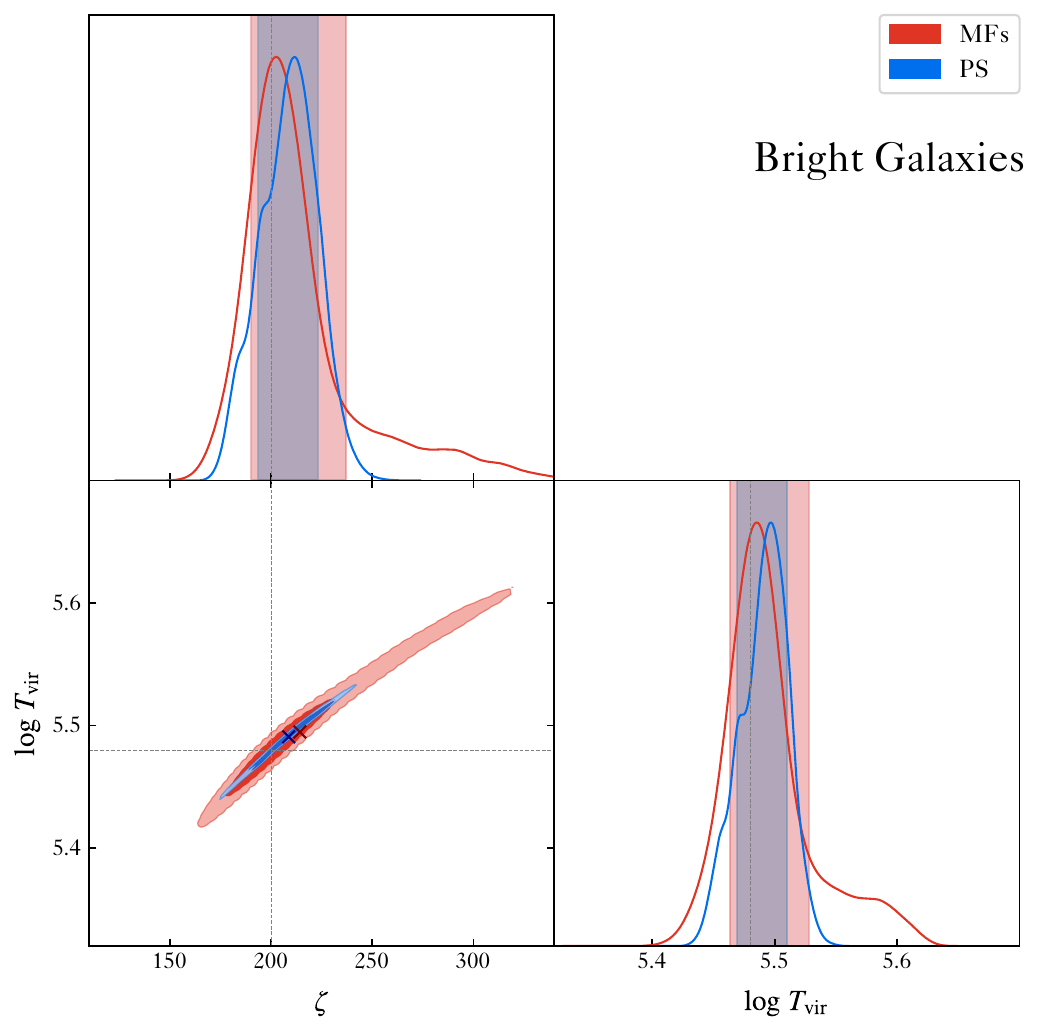}
    \caption{The posterior distribution of EoR parameters using the SKA-Low configuration with 1000\,h observation for the \fg\ (left) and \bg\ (right) model, respectively. The lower left plot in each panel shows the $1\sigma$ (dark-shaded region) and $2\sigma$ region (light-shaded region), with the results from MFs shown in red and PS shown in blue. The dashed lines indicate the fiducial parameter values. The lower-right plot and upper-left plot in each panel show the marginalized probability distribution, where $1\sigma$ intervals are shown as the shaded regions.}
    \label{fig:SKAInf}
\end{figure*}
\begin{figure*}[htbp]
    \centering
    \includegraphics[width=0.496\textwidth]{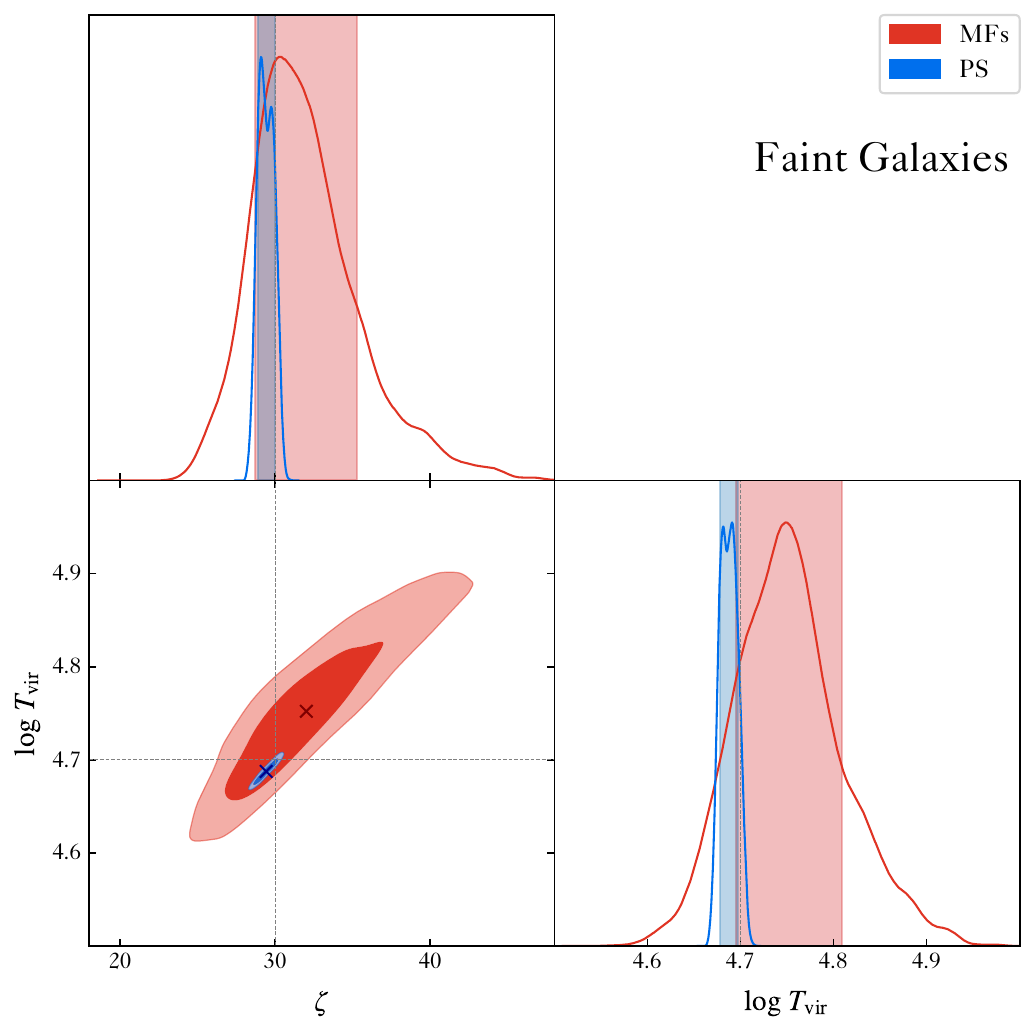}
    \includegraphics[width=0.496\textwidth]{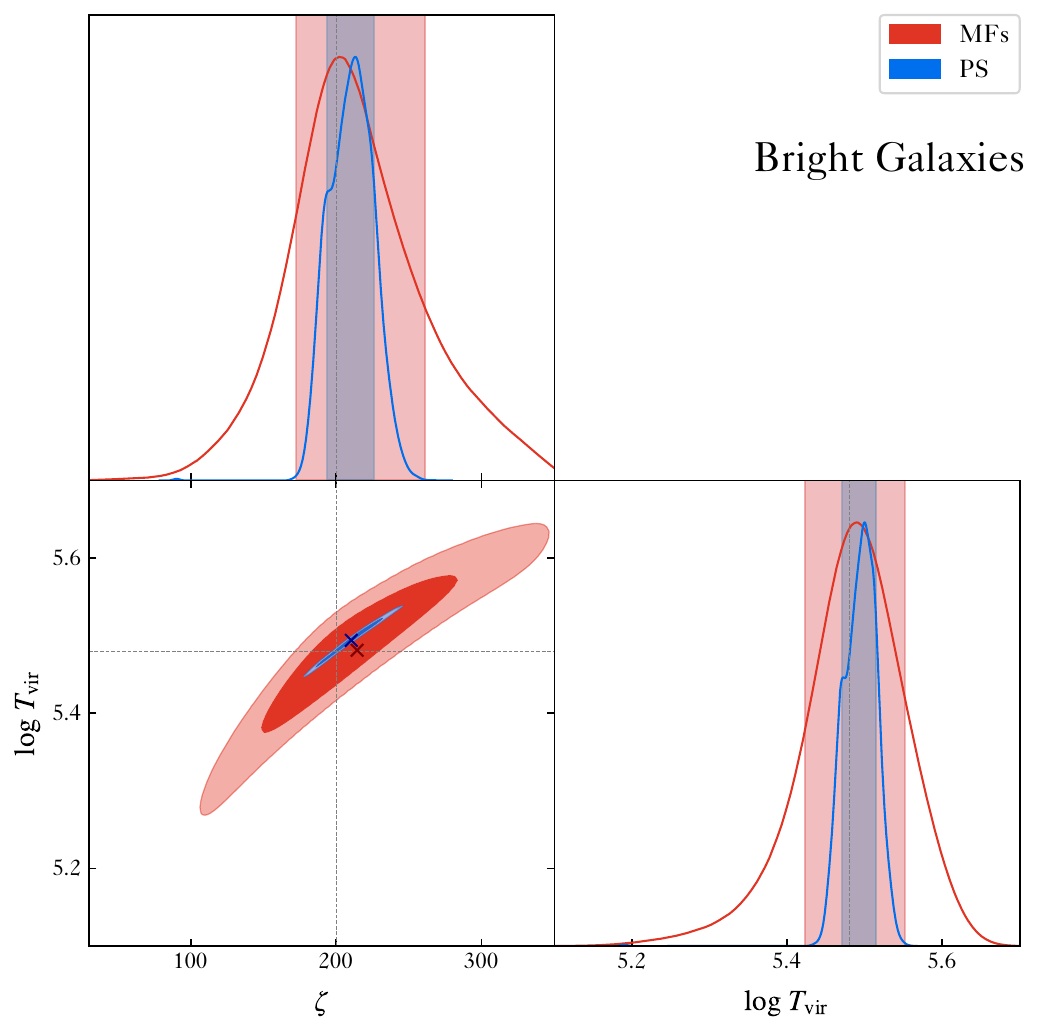}
    \caption{Same as Figure \ref{fig:SKAInf} but for the HERA-like configuration with 1000\,h observation (and with different axis scales).}
    \label{fig:HERAInf}
\end{figure*}

Our mock observations are $512\times 512\times 128$ Mpc$^3$ coeval cubes on a $256\times 256\times 64$ grid, the same as the coeval cube sampled within MCMC. Then the coeval cube and noise cube are all smoothed to the same telescope configuration before calculating MFs. 

In this proof-of-concept work, we adopt a simple Gaussian likelihood. Ignoring the correlation between any two points in MFs, the likelihood simply takes the form $\log \mathcal{L}(d\vert \theta)=-\sum_{ij}(f(\theta)_{ij}-d_{ij})^2/2\sigma_{ij}^2$, where the indices $i$ and $j$ represent the $j$-th data point in the $i$-th redshift bin, $d$ is the data to be fit, $f(\theta)$ denotes the data simulated within MCMC of parameter $\theta$ and $\sigma$ is the standard deviation of the data. In this work, we pre-calculate the $\sigma$ by forward simulation of the fiducial model 1000 times, each time varying the initial conditions and noise realizations. According to our test, the $\sigma$ starts to converge after 100 realizations, and we only exploit the MFs $\{V_1,V_2,V_3\}$ in calculating the likelihood, provided that their changes caused by varying reionization parameters are more notable than $V_{0}$. 

Two fiducial models are considered, the \fg\ model ($\zeta=30$, $\log_{10} T_{\rm vir}=4.70$) and the \bg\ model ($\zeta=200$, $\log_{10} T_{\rm vir}=5.48$) as in \citet{Greig_2017}. The parameters to be constrained are $\zeta$ and $T_{\rm vir}$. We impose a flat prior for $20 \le \zeta \le 350$ and for $4\le \rm \log_{10} T_{\rm vir} \le 6$. 
For mock observations, we again focus on the SKA-Low configuration, and also explore HERA-like scenarios for an extreme comparison, with the exact specifications listed in Table \ref{tab:telescope}.

We simplify the fitting by using comoving boxes at $z=7$ and $z=8$ instead of using lightcones. The comoving boxes with $512 \times 512 \times 128\,{\rm Mpc^3}$ boxes at $z=7$ and $z=8$ represent an approximation of splitting the lightcones between $z=7$ and $z=8$ into two sub-bands.
As discussed in Section \ref{sec:mfsevo}, the MFs between $z=7$ and $z=8$ are most sensitive to the EoR parameters, so we expect that most constraining power comes from the $z=7-8$ sub-bands.
Our \cmmc\ fitting uses 128 walkers with 100 iterations each, and the initial conditions for all MC samples are fixed. To achieve a balance between computational efficiency and inclusion of the cosmic variance, in each scenario considered in this section, we conducted ten independent runs with different initial conditions, resulting in $\sim 10^5$ samples in total.

We further investigate the parameter fitting with the 21~cm power spectrum (PS) in the same settings, providing a benchmark for the EoR information contained in MFs. The sensitivities for HERA and SKA-Low are calculated via the \cmss\ code with the lower and upper limits of $k$ corresponding to the box length and the resolution of the assumed telescope, respectively. We account for three effects that affect the sensitivity: the thermal noise of the telescope, the sampling variance of the mock observation, and the variance introduced by the different initial conditions in \cmfst\ (i.e.\ theoretical systematics). Joint constraints are also explored by simply combining the log-likelihoods, to verify if the MFs provide any information beyond the 2-point statistics.

\subsection{Results with Telescope Noise}
\label{sec:fitnofg}

We start by exploring the scenario with thermal noise but no foreground contamination, taking the fiducial parameter values of the \fg\ and \bg\ models. According to the discussion in Section~\ref{sec:mfsparams}, the former is expected for tighter constraints on reionization parameters than the latter. Our parameter constraints for SKA-Low are shown in Figure \ref{fig:SKAInf} and listed in Tables~\ref{tab:FaintInf} and \ref{tab:BrightInf}. We find that, with SKA-Low observations using MFs, both $\zeta$ and $\rm T_{\rm vir}$ can be constrained with precise measurement errors and are consistent with the fiducial values. 
For \fg\ model with SKA-Low, the constraining accuracies of $\zeta$ and $\log_{10} T_{\rm vir}$ are $(1.23\%^{+2.30\%}_{-2.33\%},0.13\%^{+0.30\%}_{-0.30\%})$ from MFs, and $(-2.13\%^{+1.70\%}_{-1.40\%},-0.30{\%}^{+0.19\%}_{-0.17\%})$ from PS. 
For \bg\ model with SKA-Low, the constraining accuracies of $\zeta$ and $\log_{10} T_{\rm vir}$ are $(7.13\%^{+11.39\%}_{-12.01\%},0.27{\%}^{+0.60\%}_{-0.57\%})$ from MFs, and $(4.36\%^{+7.17\%}_{-7.51\%},0.20\%^{+0.38\%}_{-0.36\%})$ from PS. 
Therefore, in the case of SKA-Low measurements for both \fg\ and \bg\ models, MFs demonstrate comparable (yet slightly worse) constraining powers to PS.


In comparison, the results with HERA-like configurations are 
shown in Figure~\ref{fig:HERAInf} and listed in Tables~\ref{tab:FaintInf} and \ref{tab:BrightInf}.  
For \fg\ model with HERA-like observation, the constraining accuracies of $\zeta$ and $\log_{10} T_{\rm vir}$ are $(6.70\%^{+10.90\%}_{-11.07\%},1.11{\%}^{+1.19\%}_{-1.21\%})$ from MFs, and $(-1.93\%^{+1.80\%}_{-1.87\%},-0.26{\%}^{+0.19\%}_{-0.19\%})$ from PS. 
For \bg\ model with HERA-like observation, the constraining accuracies of $\zeta$ and $\log_{10} T_{\rm vir}$ are $(7.18\%^{+23.49\%}_{-20.99\%},0.18{\%}^{+1.28\%}_{-1.08\%})$ from MFs, and $(5.02\%^{+7.66\%}_{-7.22\%},0.26{\%}^{+0.36\%}_{-0.44\%})$ from PS. We find that, with HERA observation, the constraints from MFs are much worse than those from PS, due to the lack of long baselines to resolve small-scale structures and insufficient sampling of different angular scales.

\begin{figure}[htbp]
    \centering
    \includegraphics[width=0.45\textwidth]{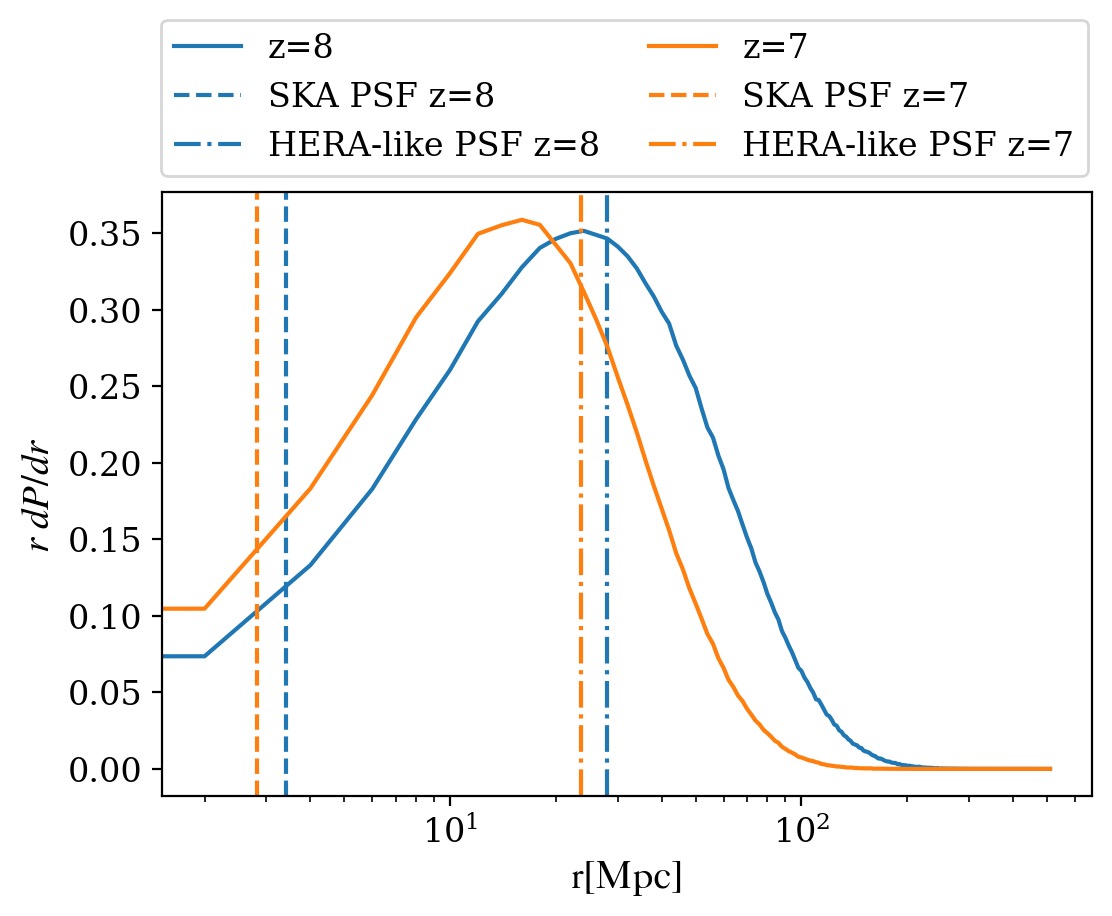}
    \caption{The bubble size distribution of the reionization field in the simulation for the \fg\ model at $z=7$ (solid yellow line) and $z=8$ (solid blue line). The FWHM of the PSF is shown for the HERA-like configuration (dot-dashed lines) and SKA-Low (dashed lines), respectively.}
    \label{fig:bsd}
\end{figure}

\begin{figure}[tbp]
    \centering
    \includegraphics[width=\linewidth]{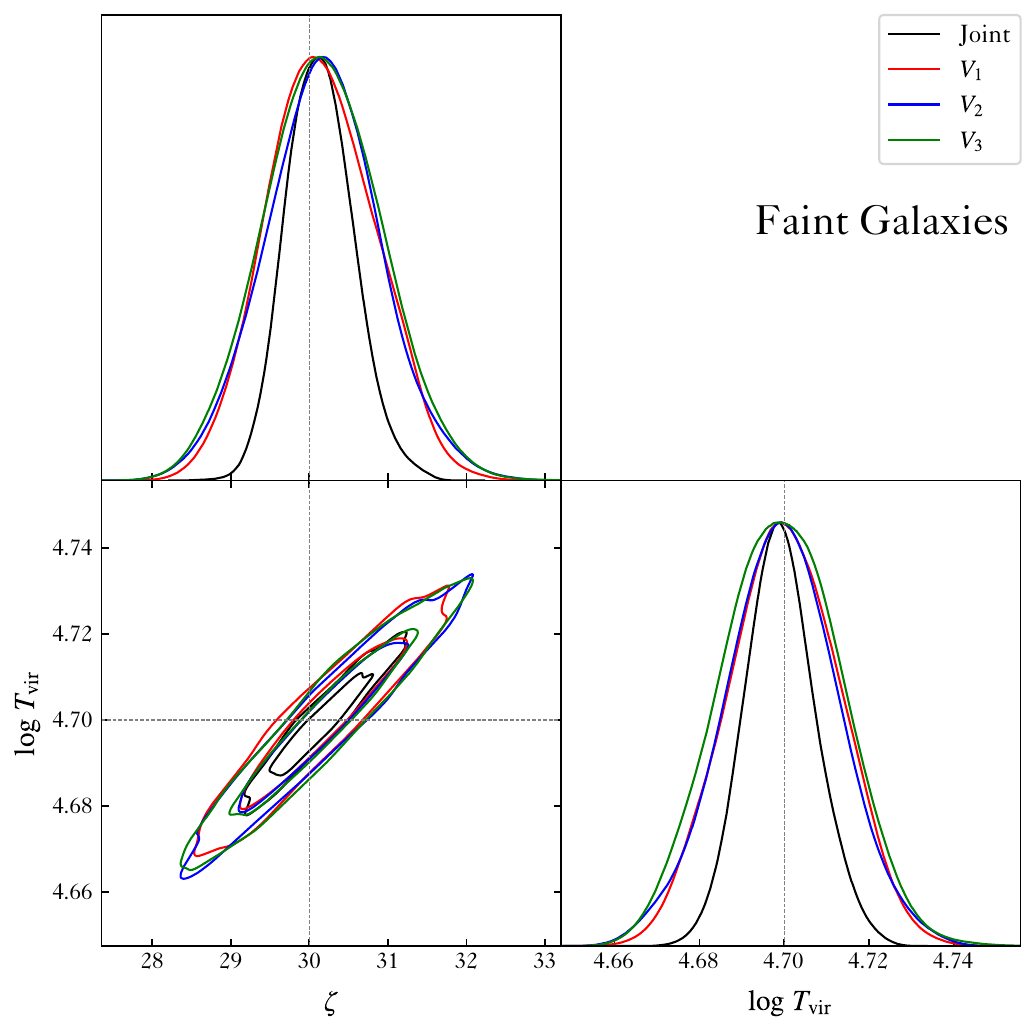}
    \caption{The posterior distribution of EoR parameters using the SKA-Low configuration with 1000\,h observation and no foreground cut, for the \fg\ model. We show the constraint using only $V_1$ (red), $V_2$ (blue), $V_3$ (green), and using all three MFs jointly (black), respectively.} 
    \label{fig:SKAfgInfSep}
\end{figure}

To further compare the HERA-like configuration with SKA-Low, we calculate the bubble size distribution (BSD) of the reionization field in our simulation for the \fg\ case and compare it with the angular resolution of the instruments shown in Figure \ref{fig:bsd}. We find that SKA-Low is capable of resolving major structures in the 21~cm field, while HERA-like configuration can only resolve the large-scale structures. Furthermore, the lack of $u$-$v$ sampling from the redundant configuration means that the thermal noise in image space is highly correlated as shown in Figure \ref{fig:Noise}, thereby further smearing the structure. The combination of the lack of long baselines and redundant layouts means that a HERA-like instrument is not ideal for topological studies of EoR.

In addition, to make a diagnostic analysis regarding which MF is more sensitive to reionization parameters than others, we perform the posterior distribution using the information of the MF $V_1$, $V_2$, and $V_3$, respectively, shown in Figure~\ref{fig:SKAfgInfSep}. We find that the inference result from each separate MF is comparable, and the joint result with all three MFs is much tighter than with any of each MF. 

In conclusion, for SKA-Low observations in the case of no foreground contamination, the MFs can be exploited as summary statistics to constrain reionization parameters with accuracies of the same order of magnitude as the power spectrum.

\begin{table*}
 \centering
 \caption{Bayesian inference with MFs and/or PS for the \fg\ model. Three scenarios are considered here --- SKA-Low observation with thermal noise, HERA-like observation with thermal noise, and SKA-Low observation with both noise and foreground cut (``FG cut''). For each scenario, we show the inference results from the MFs and from the PS, respectively, and for the SKA (noise+FG cut) scenario, the result of joint inference from both MFs and PS.}
 \begin{tabular}{ccccccccc}
 \hline\hline
    & &\multicolumn{2}{c}{SKA(Noise)} &\multicolumn{2}{c}{HERA-like(Noise)} &\multicolumn{3}{c}{SKA(Noise+FG Cut)} \\
    \cmidrule(l{.75em}r{.75em}){3-4}
    \cmidrule(l{.75em}r{.75em}){5-6}
    \cmidrule(l{.75em}r{.75em}){7-9}
 Parameter & True value & {MFs} & {PS} & {MFs} &  {PS} & {MFs} & {PS}&{Joint}\\
   \hline
   $\log _ { 10 } \left( T_ { \rm vir }/{\rm K}\right)$ & 4.700 & $4.706^{+0.014}_{-0.014}$&  $4.686^{+0.009}_{-0.008}$ &  $4.752^{+0.056}_{-0.057} $ &  $4.688^{+0.009}_{-0.009}$&  $4.718^{+0.022}_{-0.024}$ &  $4.689^{+0.010}_{-0.009}$&
   $4.699^{+0.008}_{-0.009}$\\
   \hline
   $\zeta$ & 30 & $30.37^{+0.69}_{-0.70}$ & $29.36^{+0.51}_{-0.42}$ & $32.01^{+3.27}_{-3.32}$ & $29.42^{+0.56}_{-0.54}$ & $31.14^{+1.06}_{-1.16}$& $29.46^{+0.55}_{-0.54}$&
   $30.05^{+0.46}_{-0.49}$\\

 \hline\hline
 \end{tabular}
 \label{tab:FaintInf}
\end{table*}

\begin{table*}
 \centering
 \caption{Same as Table \ref{tab:FaintInf} but for the \bg \ model.}
 \begin{tabular}{ccccccccc}
 \hline\hline
    & &\multicolumn{2}{c}{SKA(Noise)} &\multicolumn{2}{c}{HERA-like(Noise)} &\multicolumn{3}{c}{SKA(Noise+FG Cut)} \\
    \cmidrule(l{.75em}r{.75em}){3-4}
    \cmidrule(l{.75em}r{.75em}){5-6}
    \cmidrule(l{.75em}r{.75em}){7-9}
 Parameter & True value & {MFs} & {PS} & {MFs} &  {PS} & {MFs} & {PS} &{Joint}\\
   \hline
   $\log _ { 10 } \left( T_ { \rm vir }/{\rm K}\right)$ & 5.480 & $5.495^{+0.033}_{-0.031}  $& $5.491^{+0.021}_{-0.020}$ & $5.481^{+0.070}_{-0.059} $ & $5.494^{+0.020}_{-0.024}$& $5.495^{+0.044}_{-0.043}$ & $5.475^{+0.028}_{-0.042}$&
   $5.477^{+0.024}_{-0.027}$\\
   \hline
   $\zeta$ & 200 & $214.26^{+22.78}_{-24.02}$ & $208.72^{+14.34}_{-15.02}$ & $214.36^{+46.97}_{-41.97} $ & $210.40^{+15.33}_{-14.43}$ & $214.65^{+30.64}_{-30.33}$& $198.32^{+18.88}_{-26.41}$&
   $199.83^{+16.14}_{-17.84}$\\

 \hline\hline
 \end{tabular}
 \label{tab:BrightInf}
\end{table*}

\subsection{Results with Noise and Foreground Avoidance}
\label{sec:fitfg}
\begin{figure*}[htbp]
    \centering
    \includegraphics[width=0.496\textwidth]{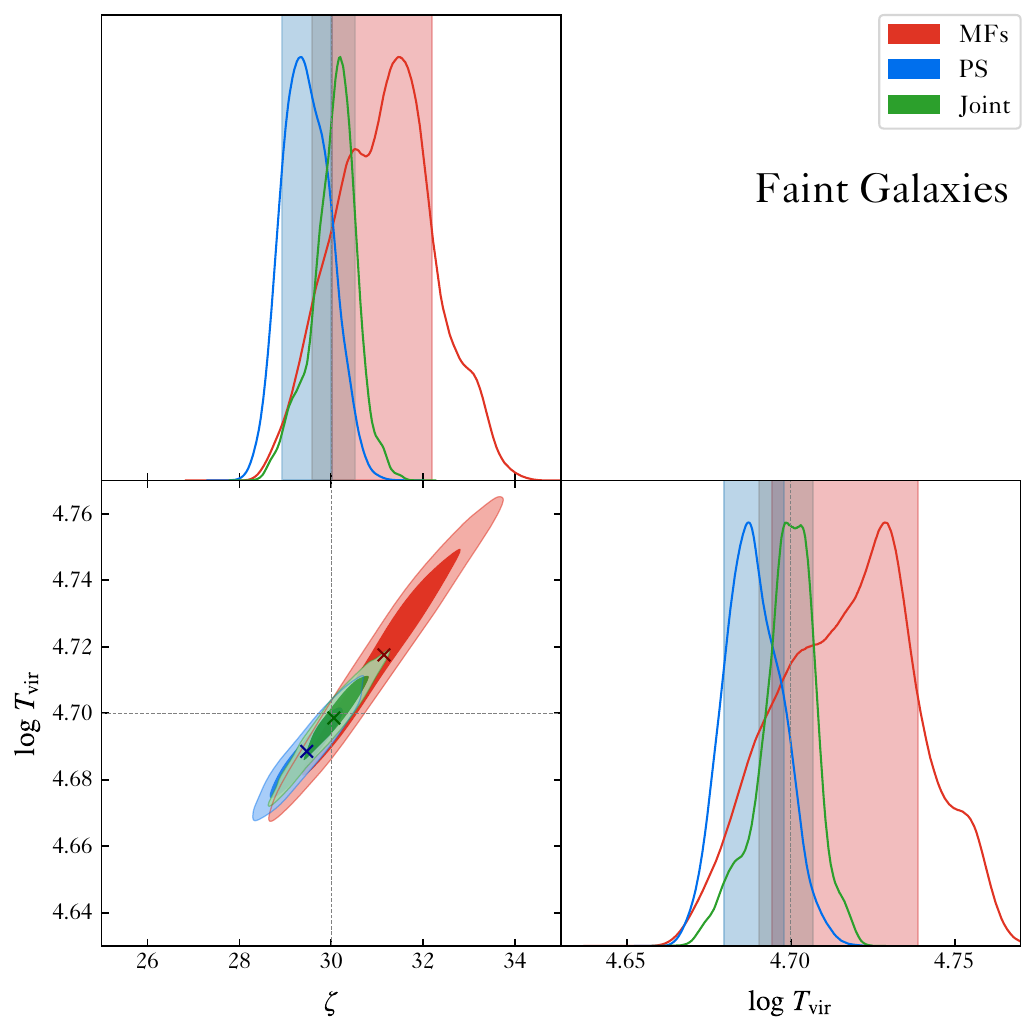}
    \includegraphics[width=0.496\textwidth]{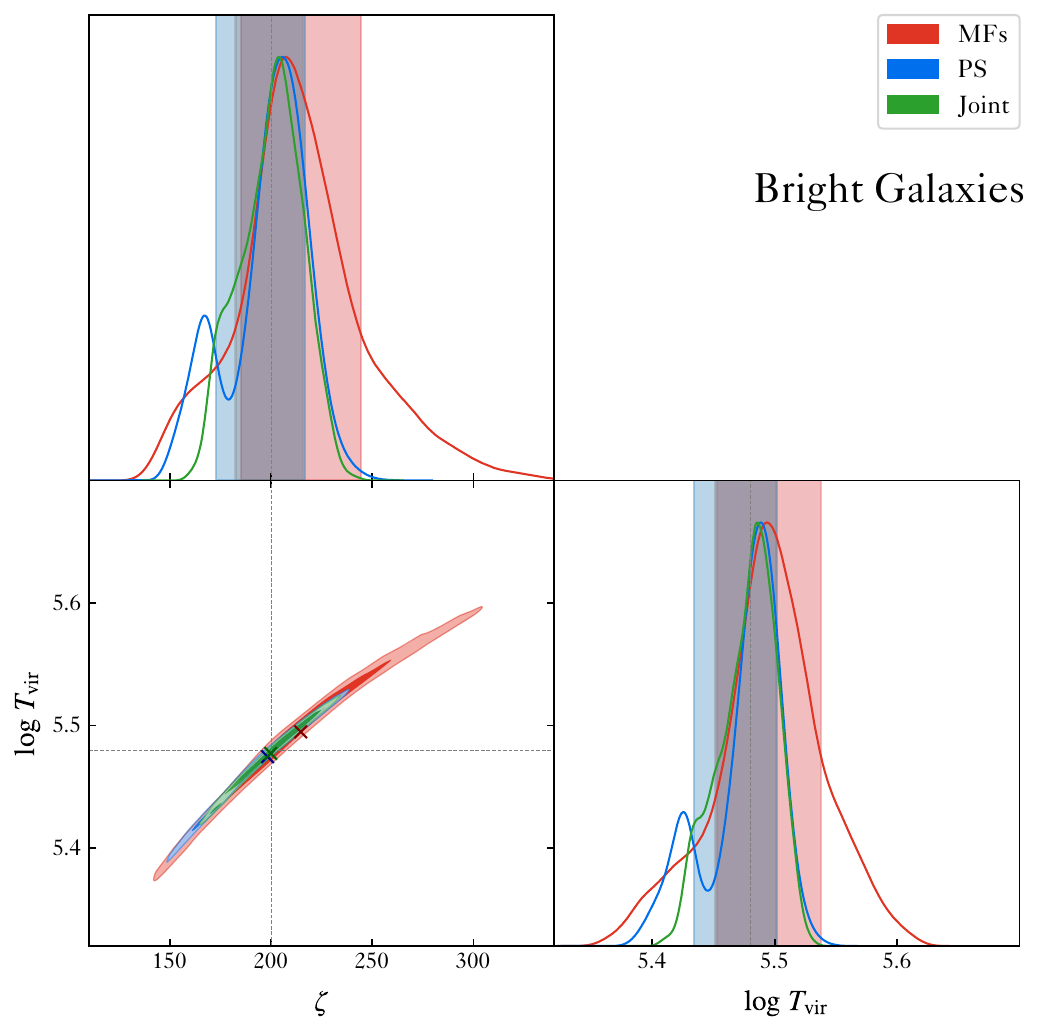}
    \caption{Same as Figure \ref{fig:SKAInf} using the SKA-Low configuration but the foreground cut is applied herein. We also show the joint constraint (green region).}
    \label{fig:SKAfgInf}
\end{figure*}
Now we test the robustness of the constraining power of MFs in the presence of the foregrounds, following the foreground avoidance recipe described in Section \ref{sec:fgwedge}.
In Figure~\ref{fig:SKAfgInf}, we show the results with the foreground avoidance technique. Here we only consider SKA-Low, because, for the HERA mock observation, the information along LoS is almost lost, as we discussed in Section~\ref{sec:mfsconf}. We adopt the foreground wedge slope $m=0.7$ herein, the more realistic scenario discussed in Sections~\ref{sec:fgwedge} and \ref{sec:mfsconf}.

Assuming the foreground avoidance, for \fg\ model with SKA-Low observation, the constraining accuracies of $\zeta$ and $\log_{10} T_{\rm vir}$ are  $(3.80\%^{+3.53\%}_{-3.87\%},-0.38{\%}^{+0.47\%}_{-0.51\%})$ from MFs, and $(-1.80\%^{+1.83\%}_{-1.80\%},-0.23\%^{+0.21\%}_{-0.18\%})$ from PS.  
For \bg\ model with SKA-Low observation, the constraining accuracies of $\zeta$ and $\log_{10} T_{\rm vir}$ are $(7.33\%^{+15.32\%}_{-15.17\%},0.27\%^{+0.80\%}_{-0.78\%})$ from MFs, and $(-0.84\%^{+9.44\%}_{-13.21\%},-0.09\%^{+0.51\%}_{-0.77\%})$ from PS. This indicates that after removing the information underneath the foreground wedge, 
the constraining powers of both MFs and PS are degraded due to loss of information.
Comparing the change in the error bars of the parameter fittings, MFs lose more information in the \fg\ case than PS. On the other hand, in the \bg\ case, the fractional changes in error bars are similar for MFs and PS. 
The larger fractional decrease in constraining power for MFs in the \fg\ model is because the MFs are more sensitive to the topology of the reionization process, which has more impact in the \fg\ model that generates a larger population of small ionized bubbles.
Nevertheless, while the fractional signal loss is larger in the \fg\ case, the absolute constraining power of MFs is stronger in the \fg\ case than in the \bg\ case. 

We find that results for both PS and MFs show slight bi-modality in the 1D posterior distribution,
which is also reported in \citet{watkinson2021epoch}. It hints towards the fact that the fitting is susceptible to parameter degeneracy. As MFs and PS probe different information about the reionization process, we expect that the joint analysis can break the degeneracy and give tighter constraints, which we will explore in Section~\ref{sec:fitjoint}.

In conclusion, we find that the MFs are robust against reasonable levels of thermal noise and foreground cut and can be used to constrain EoR parameters. For SKA-Low observation, MFs have comparable constraining powers to PS, making them a set of promising summary statistics beyond the two-point correlation. 

\subsection{Joint Constraints using MFs and PS}
\label{sec:fitjoint}
Now, we further explore the constraining power of MFs, by studying the joint analysis using MFs and PS in the parameter fitting. 
The 2D posterior is shown in Figure~\ref{fig:SKAfgInf} with the fitting results listed in the rightmost columns of Tables~\ref{tab:FaintInf} and \ref{tab:BrightInf}.


The joint analysis gives tighter constraints for both \bg\ and \fg\ cases than the constraints using MFs or PS alone. In particular, for the \fg\ case, compared to the results from only PS, joint fittings improve about $\sim 10\%$ accuracies to both $\zeta$ and $\log_{10} T_{\rm vir}$ measurements. For the \bg\ case, the improvements on the parameter accuracies increase to $\sim 20\%-30\%$.

Furthermore, as shown in Figure \ref{fig:SKAfgInf}, the joint constraints also break the parameter degeneracy. In the \fg\ case, the 1$\sigma$ credible region from the MFs and that from PS are slightly misaligned, with the former significantly larger. As a result, the best fit of the joint analysis is more precise about the true values, with the joint credible region further shrunk from the PS credible region. 
In the \bg\ case, the credible region from PS finds a local (but not global) likelihood maximum at the lower values of the reionization parameters than the true values, while the credible region from the MFs is very elongated. In comparison, the joint analysis has only one maximum of likelihood, and the joint credible region is much smaller than that of the MFs or PS. 
In general, regarding the improvement of joint analysis upon PS, we find that the improvement is more significant in the \bg\ case than in the \fg\ case.


We conclude from these results that the MFs contain complementary information about the EoR parameters to the PS. 

\subsection{Caveats}
In this subsection, we discuss several points that may affect the results. 

First of all, the morphological structure in the 21~cm brightness temperature field may be oversimplified in this work because the field is simulated using \cmfst\ which is based on the excursion set model of reionization. Realistic astrophysical processes, e.g. inclusion of inhomogeneous recombination (e.g. \citealt{2020MNRAS.491.1600M}), might result in a more sophisticated morphological structure, which might enhance the constraining power of MFs.

In addition, several observational effects such as redshift-space distortion (e.g. \citealt{2012MNRAS.422..926M}) and lightcone effect (e.g. \citealt{2012MNRAS.424.1877D}), which are neglected in this work, can introduce large anisotropic effects in the 21~cm signal. Since the foreground wedge affects the sampling of the anisotropic modes along the LoS, how MFs are sensitive to reionization parameters is subject to further tests. 


We also implicitly assume in this work that the foreground wedge corresponds to the angular scale that is the instrument FoV of the primary beam and neglects the effect of sidelobes. This assumption requires sufficient mitigation of the wide-field effect in 21~cm observations. The mitigation of the wide-field effect is performed in data calibration and imaging, which is beyond the scope of this paper. 

Moreover, even assuming perfect wide-field mitigation, residual foregrounds can still significantly contaminate the data \citep{2024MNRAS.tmp..283B}. Investigating the impact of residual foreground contamination and incorporating it into the inference pipeline requires accurate and efficient modeling of the foreground emission, which is left to future work.


Finally, our posterior inference adopts the standard MCMC algorithm with a simple Gaussian likelihood and, in particular, neglects the correlation between any threshold $\delta T_b/\sigma_{T_b}$ bins and between any redshift bins in the MFs. This treatment is sufficient for our forecast study, but it will be necessary for future realistic analysis to include the complete modeling of the full covariance matrix in the MCMC analysis \citep{2023MNRAS.524.4239P} or perform the simulation-based inference \citep[SBI;][]{zhao2021,Zhao2022} using the MFs.

\section{Summary}\label{sec:sum}
In this paper, we explore the viability of using MFs as summary statistics for 21~cm observations during the EoR. 

First, we test the robustness of MFs in the presence of observational effects by generating mock observations that account for the effects of thermal noise, synthesized beam, and foreground avoidance. The mock observations are simulated using the configurations of SKA-Low, as well as a redundant array layout similar to HERA. 


We find that the synthesized beam of radio interferometers smooths the 21~cm brightness temperature field, altering the structure of the MFs. The Gaussian thermal noise obscures the underlying structure of reionization. Both effects result in the loss of information in the non-Gaussian reionization process.

For redundant array configurations, the lack of long baselines, which results in insufficient sampling in different angular scales, takes out the structures of the MFs in the observed 21~cm signal. As a result, HERA-like configurations are not ideal for constraining EoR models.

For the SKA-Low array, on the other hand, sufficient sensitivity and baseline coverage help retain the information in the MFs. Assuming a deep total integration time of 1000 hours, the topology of reionization can be seen in the observed signal with its clear evolution, and as such, MFs can be used to distinguish different stages of reionization as well as different models.

It is expected that foreground avoidance causes further loss of information in the MFs, making them more Gaussian. However, assuming that the foreground wedge corresponds to the primary beam of the instrument, we find that there is still enough information in the observation window for MFs to capture the topology of EoR.

Next, we forecast the prospects of constraining EoR parameters with MFs. For this purpose, we build an MCMC parameter-fitting framework that uses forward modeling. 
We find that for deep observations using SKA-Low, MFs can be used to provide unbiased estimations of the reionization parameters, with the forecast sensitivity comparable to the PS. 
For the HERA-like configurations, nevertheless, the constraint accuracies of the reionization parameters using MFs are about ten times worse than using the PS. 

When the foreground cut is applied, the constraint accuracies of the reionization parameters with the MFs are degraded, as we demonstrate with SKA-Low observations. 
The presence of foregrounds leads to more information loss in the MFs compared to the PS because the MFs are more sensitive to the topology of the reionization fields.

Jointly fitting the parameters using both MFs and PS, we find that the joint constraints are tighter than each separate fitting, which implies that the MFs contain complementary information to the two-point statistics. The improvements in the constraint accuracies depend on the reionization model. Compared to the accuracy using only PS, the accuracies in the reionization parameters are improved with $\sim 30\%$ ($10\%$) in the \bg\ model (\fg\ model). 


Our investigation highlights the prospect of constraining the reionization parameters by exploiting the MFs as complementary summary statistics to the PS. With the upcoming SKA-Low that will generate high-fidelity 21~cm images with a large dynamical range, the full topological information in the 21~cm image cube will be crucial in understanding the reionization and the driving force behind it.

\section*{acknowledgments}
This work is supported by the National SKA Program of China (grant No.~2020SKA0110401), NSFC (grant No.~11821303, 12361141814), and the Chinese Academy of Sciences grant ZDKYYQ20200008. 
ZC's research is funded by a UKRI Future Leaders Fellowship [grant MR/X005399/1; PI: Alkistis Pourtsidou]. 
KD thanks Xiaosheng Zhao, Ce Sui, Bohua Li, and Meng Zhou for inspiring discussions and help. 
We acknowledge the Tsinghua Astrophysics High-Performance Computing platform at Tsinghua University for providing computational and data storage resources that have contributed to the research results reported within this paper.


\vspace{5mm}

\software{\cmfst\citep{21cmFAST_1.0,21cmFAST_3.0}, \cmmc\citep{Greig_2015,Greig_2017,Greig_2018}, \cmss\citep{Pober_2013,Pober_2014}, \cmtool\citep{Giri_2020}, \textsc{\small OSCAR}, \textsc{\small HERA-sim}.}

\bibliography{reference}

\end{document}